\documentclass[12pt, draftclsnofoot, onecolumn]{IEEEtran}
\pdfoutput=1
\normalsize

\usepackage{cite}
\usepackage{amsbsy}
\usepackage{amsmath,amssymb,amsfonts}
\usepackage{algorithmic}
\usepackage{graphicx}
\usepackage{textcomp}
\usepackage{graphicx}                                            
\usepackage{multicol}         
\usepackage{caption}
\usepackage{amssymb,amsmath,dsfont}
\usepackage{inputenc}
\usepackage{bbm}
\usepackage{comment}
\usepackage{cite}
\usepackage{color}
\usepackage{multirow}
\usepackage{subcaption}
\usepackage{array}
\usepackage{algorithm}
\usepackage{color,soul}
\usepackage{blindtext}
\usepackage{bm}
\usepackage{dsfont}
\usepackage[sort&compress,numbers]{natbib}
\usepackage{mdframed}
\setcitestyle{square}

\makeatletter
\newcommand{\mathleft}{\@fleqntrue\@mathmargin0pt}
\newcommand{\mathcenter}{\@fleqnfalse}
\makeatother

\newtheorem{proposition}{Proposition}
\newtheorem{assumption}{Assumption}

\newtheorem{proof}{Proof}
\newtheorem{remark}{Remark}

\usepackage[symbol]{footmisc}

\begin{document}
\title{Joint data rate and EMF exposure analysis in user-centric cell-free massive MIMO networks}

\author{Charles~Wiame,~\IEEEmembership{Member,~IEEE}, Claude~Oestges,~\IEEEmembership{Fellow,~IEEE}, and~Luc~Vandendorpe,~\IEEEmembership{Fellow,~IEEE}
\thanks{This work has been submitted to the IEEE for possible publication. Copyright may be transferred without notice, after which this version may no longer be accessible.}}

\maketitle

\begin{abstract}
The objective of this study is to analyze the statistics of the data rate and of the incident power density (IPD) in user-centric cell-free networks (UCCFNs). To this purpose, our analysis proposes a number of performance metrics derived using stochastic geometry (SG). On the one hand, the first moments and the marginal distribution of the IPD are calculated. On the other hand, bounds on the joint distributions of rate and IPD are provided for two scenarios: when it is relevant to obtain IPD values above a given threshold (for energy harvesting purposes), and when these values should instead remain below the threshold (for public health reasons). In addition to deriving these metrics, this work incorporates features related to UCCFNs which are new in SG models: a power allocation based on collective channel statistics, as well as the presence of potential overlaps between adjacent clusters. Our numerical results illustrate the achievable trade-offs between the rate and IPD performance. For the considered system, these results also highlight the existence of an optimal node density maximizing the joint distributions.
\end{abstract}


\IEEEpeerreviewmaketitle

\section{Introduction}


During the past decade, cell-free (CF) networks have progressively been subject to numerous publications in the telecommunication literature. Several benefits of this architecture can be highlighted when comparing it with traditional cellular networks: improved interference management capabilities, increased SNR values (thanks to coherent transmissions), and a more uniform coverage \cite{Intro_2}. Thanks to these assets, this network paradigm is considered as a promising solution for beyond 5G mobile communications \cite{Intro_1}. 

\begin{figure}[h!]
  \centering
  \belowcaptionskip = -5pt
  \begin{subfigure}{0.5\textwidth}
    \includegraphics[trim={0.3cm 0 0 0},scale=0.207]{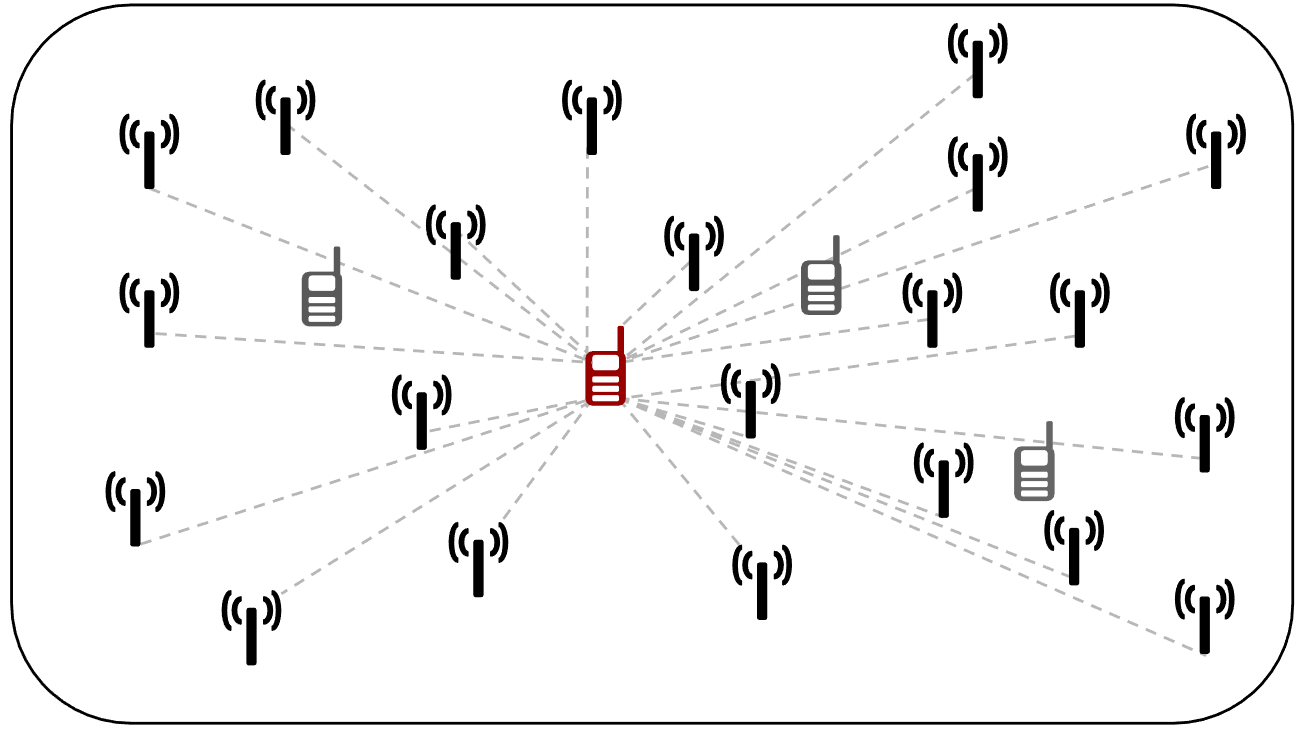}
    \caption{Original CF network where each user can \\potentially be served by any radio head.} 
    \label{all_vs_centric:a}
  \end{subfigure}%
  \hfill
  \begin{subfigure}{0.5\textwidth}
  \centering
    \includegraphics[trim={0.4cm 0 0 0},scale=0.207]{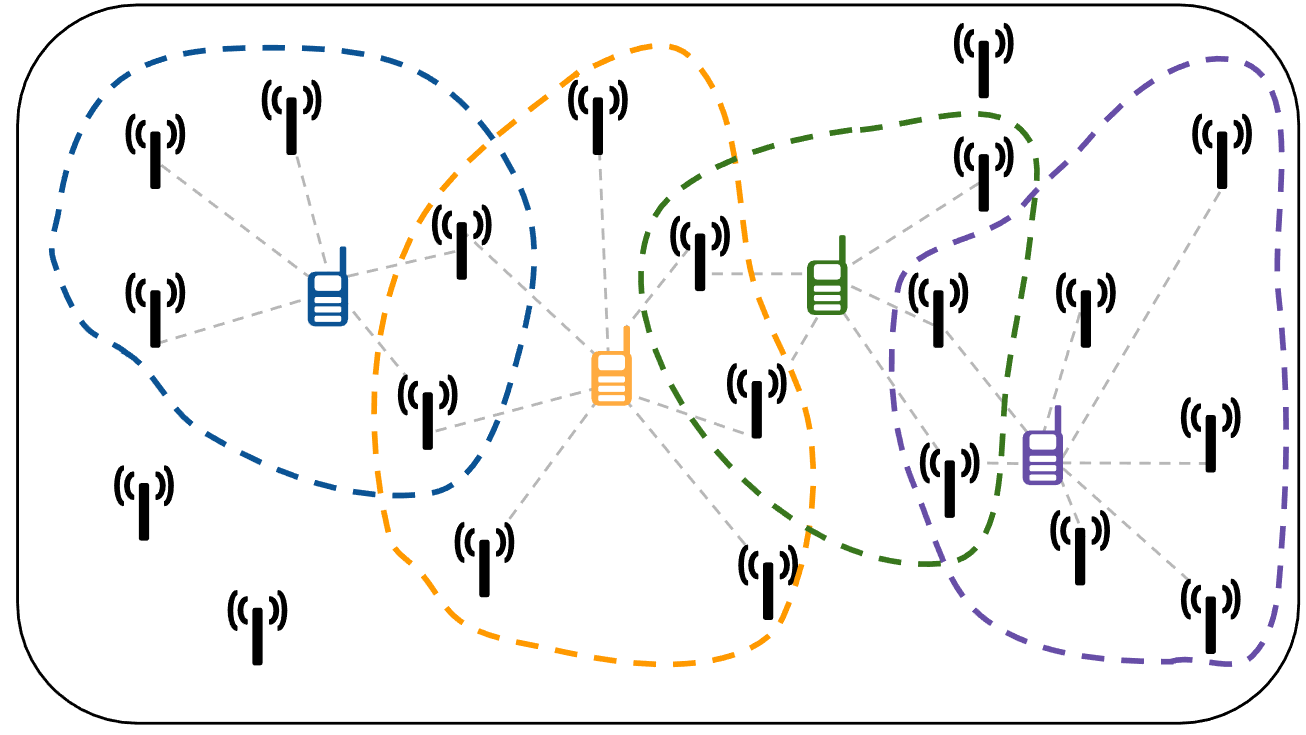}
    \caption{UCCFN where clusters are defined around every user.\\} \label{all_vs_centric:b}
  \end{subfigure}
  \vfill
  \caption{Cell-free versus user-centric cell-free networks.}
\end{figure}

In addition to the original CF structure, one will note the existence of a \textit{user-centric} version \cite{8000355}. In the first works considering the original CF architecture \cite{7227028}, each user is simultaneously served by all radio heads (Figure \ref{all_vs_centric:a}). By contrast, user-centric cell-free networks (UCCFNs) restrict the serving access points to spatial subsets defined around every user (Figure \ref{all_vs_centric:b}). In the rest of this work, these subsets will be named \textit{cooperative clusters}. Such cluster structure is motivated by the scalability and reduced complexity obtained thanks to the access point selection \cite{9064545}. In both cases, the signal transmitted by each radio head is composed of a superposition of precoded signals intended for the UEs served by this radio head. In the original CF architecture, this superposition hence includes all UEs in the network. For UCCFNs, it can include one or more users if the radio head is selected to be part of several clusters \cite{Intro_2}, which explains the overlaps in Figure \ref{all_vs_centric:b}. The system model of this study focuses on this user-centric adaptation. 

Most existing works related to CF networks have proposed a performance analysis relying on data-related metrics: coverage probability, spectral efficiency, etc. However, few publications have studied the statistics of the incident power density (IPD) affecting mobile users within this architecture. This quantity can be of particular importance in modern scenarios: either for wireless power transfer (WPT) purposes, or for public health concerns. Power density analyses in existing CF-WPT works are partial: either no statistics are derived, or only the first moments of the incident power are computed. This paper aims at bridging the resulting gap by deriving the individual and joint distributions of the IPD and rate in CF networks. In order to calculate these distributions, the analytical framework of this study relies on stochastic geometry (SG). This branch of spatial statistics enables to abstract network nodes (users and base stations) as point processes. Using the properties of these processes, metrics of interest can be analytically computed, and averaged over all potential nodes locations \cite{Baccelli}.

\subsection{Related works}

This section is structured as follows: first, existing SG studies related to CF networks are detailed. Second, an overview of SG works analyzing the IPD in general networks is provided.


\subsubsection{Studies related to CF networks}

the analysis in \cite{8972478} describes a CF network generated via a Poisson point process (PPP). The spectral efficiency of a typical user equipment (UE) is derived. The authors of \cite{8379438} establish conditions for channel hardening in CF systems. In \cite{9353695}, an optimization problem is formulated to maximize the energy efficiency of a CF network as function of the node densities and pilot reuse factor. Another optimization problem is proposed in \cite{9042221}, aiming at minimizing the total transmit power of a CF network. The presented model considers coherent and non-coherent transmissions, as well as spatially correlated channels. The authors of \cite{9894287} investigate the impact of fronthaul links of finite capacity on the performance. It is shown that the node deployment should be more distributed when the quality of the channel state information decreases. The model in \cite{9394765} considers a time division duplexing CF system, whose frame length is optimized to limit Doppler shift effects. The framework of \cite{8974591} analyzes a CF network featuring mobile edge computing functionalities: multiples user types are introduced, each category performing computing tasks with a given time requirement. The authors of \cite{9838432} investigate optimality conditions for the interference to be approximated as noise in CF networks. Finally, \cite{9629311} investigates the performance of CF systems with millimeter wave fronthaul.


\subsubsection{Studies related to SG-IPD}

\cite{SG-SWIPT1} derives the joint coverage-IPD distribution of a network including directional beamforming, as well as line-of-sight (LoS) and non-line-of-sight (nLoS) links. This joint distribution is also derived in \cite{SG-SWIPT2} in the case of a multiple input/multiple output (MIMO) network. Another MIMO system, employing maximum ratio transmission (MRT) is also considered in \cite{SG-SWIPT3}. The work in \cite{SG-SWIPT4} separately investigates simultaneous information and power transfer (SWIPT) in a K-tier network. The authors of \cite{SG-SWIPT5} introduce a cooperative SWIPT protocol applied in the framework of non orthogonal multiple access (NOMA) transmissions. In \cite{SG-SWIPT6}, the node density of an ad-hoc network is optimized to maximize the area energy harvested, given a constraint on the capacity. In \cite{SG-SWIPT7}, the deployment of a SWIPT-MIMO network is considered in an indoor environment. The framework introduced in \cite{7268782} aims at analyzing a relay selection problem in a network with wireless powered nodes. At the time of writing this paper, the analysis published in \cite{9045726} and \cite{9201540} constitutes the only SG work studying SWIPT in a CF architecture. The authors of this study derive the mean and the variance of the energy harvested by the UE and characterize the steady-state probability of the UE battery to be charged.


Finally, a few works analyze the IPD outside the WPT context. These studies consider the received power as an electromagnetic field (EMF) exposure that should be monitored for safety reasons. In \cite{9462948}, the IPD is analyzed in a homogeneous PPP and compared with experimental measurements conducted in an urban environment. The authors of \cite{9232290} perform a fitting of a 5G massive MIMO antenna pattern using a multimodal normal law. The resulting distribution is then incorporated in a SG analysis. The framework of \cite{9768997} consider a downlink max-min fairness power control protocol. The impact of this protocol on the EMF exposure in a MIMO network is quantified. Finally, the model of \cite{9511258} studies the IPD in scenarios including antennas operating at both mmWave and sub-6 GHz frequencies.

\subsection{Contributions}

On the basis of the aforementioned works, the main findings of this study can be summarized as follows: 

\begin{enumerate}

\item \textit{Performance metrics:} this paper provides a comprehensive analysis of the rate and IPD statistics in user-centric cell-free networks (UCCFNs). To this purpose, our analytical framework enables to derive the following metrics: 

\begin{itemize}
    \item the first moments of the total IPD measured at the UE;
    \item the marginal distributions of the data rate and of the IPD; 
    \item lower and upper bounds on the joint distribution of the rate and IPD. Two versions of the metric are considered depending on the role of the IPD in the scenario of interest. If the network model involves SWIPT aspects, the UE might be able to harvest the ambient energy. In that case, the network planner might simultaneously target high coverage and IPD values. The first version of the joint distribution (originally introduced in \cite{JCCDF}) hence requires the two quantities to be above given thresholds. By contrast, if the network does not include SWIPT aspects, the network planner might instead desire a low exposure (for public health reasons) while meeting quality of service demands. The second version of distribution therefore requires the rate to be greater than a first threshold, and the IPD lower than a second threshold. 
\end{itemize}
To the best author's knowledge, this work is the first analysis introducing the IPD distribution, as well as the joint metrics for UCCFNs. 

\item \textit{System modeling:} this study includes system features related to UCCFNs which, to our best knowledge, have not been included so far in SG models:

\begin{itemize}
    \item the power allocation is based on collective channel statistics: given cluster serving a UE, the knowledge of the channels associated to all the RRHs of that cluster is employed to perform the allocation. The distribution guarantees a fixed transmit power budget per UE. This budget is then distributed among the remote radio heads (RRHs) serving the UE. The fraction of power allocated to each of these RRHs is computed based on their relative channel gains.
    \item a new mathematical framework is provided to take into account the cluster overlaps (Figure \ref{all_vs_centric:b}) in the case of distance-based association policies. The correlations between the useful and interfering signals resulting from these overlaps are captured in our mathematical developments. The impact of these correlations on the performance is then quantified by our numerical simultations.
    
\end{itemize}

\item \textit{Performance analysis:} On the basis of the developed framework, the numerical results of this study enable to illustrate:
\begin{itemize}
    \item the achievable trade-offs in the rate and IPD requirements;
    \item for the considered system model, the existence of an optimal node density maximizing the joint data rate and IPD distributions.
\end{itemize}

\end{enumerate}

\subsection{Organization of the paper}
The rest of this paper is organized as follows: the system model is described in Section II. The analytical results derived using SG are presented in Section III. Finally, the numerical results and the conclusion are respectively detailed in Sections IV and V. 

\subsection{Notations}
In the next sections, $[\mathbf{a}_p]_{p \in \mathcal{P}}$ represents the concatenation of vectors $\mathbf{a}_p$ whose indexes $p$ belong to the set $\mathcal{P}$; $\mathcal{D}(\mathbf{x},R)$ is the disk of radius $R$ centered around $\mathbf{x} \in \mathbb{R}^2$; $|\mathcal{C}|$ denotes the cardinal number of set $\mathcal{C}$; $j$ represents the imaginary unit; $\Gamma(\cdot)$ denotes the Gamma function; $\mathbb{E}_A\big[F(A)\big]$ denotes the expectation with respect to random variable $A$ of an expression $F(A)$.

\section{System Model}
\label{sect:system}

\subsection{Network topology}
We consider the downlink of a UCCFN, as represented in Figure \ref{all_vs_centric:b}. The RRH locations are modelled by means of a homogeneous PPP $\Psi_R \subset \mathbb{R}^2$ of intensity $\lambda_R$. Given one realization of this point process, we denote $\mathbf{x}_i$, the Cartesian coordinates of RRH $i \in \Psi_R$. All RRHs are equipped with \textit{M} antennas and are connected via fronthaul links to a baseband processing unit (BBU). Users equipments (UEs) are distributed using a second homogeneous PPP $\Psi_U$, of intensity $\lambda_U$\footnote[2]{In the case of a cell-free network, we will usually consider that $\lambda_U \ll \lambda_R$ \cite{Intro_2}.}. This point process is deployed over an area $\mathcal{A} \subset \mathbb{R}^2$ conditioned on the prior realization of $\Psi_R$. This area takes into account small exclusion zones around each RRH and is defined as 
\begin{equation}
\label{area_A}
 \mathcal{A} = \mathbb{R}^2 \setminus \Big( \cup_{i \in \Psi_R} \mathcal{D}(\mathbf{x}_i,r_0) \Big)
\end{equation}
where $r_0 = 1 \; m$ is an exclusion radius enabling to avoid singularities in the path loss (see next subsection). Each UE is equipped with a single antenna.

\subsection{Channel model}

The channel vector $\mathbf{g}_{iu} \in \mathbb{C}^{M \times 1}$ between RRH $i$ and user equipment $u$ is given by 
\begin{equation}
    \label{PL_expression}
    \mathbf{g}_{iu} = \mathbf{h}_{iu} \kappa^{-1/2} r_{iu}^{-\alpha/2}.
\end{equation}
In this definition, the following elements have been introduced:

\begin{itemize}
    \item $\mathbf{h}_{iu} \in \mathbb{C}^{M \times 1}$, a vector containing the fading coefficients $h_{iuk}$ (with $k=1,\hdots,M$). These coefficients are assumed to be independently distributed. We consider Rayleigh fading, therefore $|\mathbf{h}_{iuk}|^2 \sim \exp(1)$.
    \item $\kappa r_{iu}^{\alpha}$ is the path loss where 
    \begin{itemize}
        \item $\kappa = (4\pi f / c)^2$, is the reference path loss at a distance of $1$ meter, where $f$ is the carrier frequency and $c$ the speed of light;
        \item $\alpha > 2$ is the path loss exponent;
        \item $r_{iu}$ is the Euclidean distance between $i$ and $u$. Since $\Psi_U$ is defined over the area given by \eqref{area_A}, potential singularities in \eqref{PL_expression} are circumvented.
    \end{itemize}
\end{itemize}

For the sake of mathematical tractability, shadowing effects are not modelled in this paper. 

\subsection{Association policy}

The cooperative clusters of the UCCFN are here distance based: every UE $u$ is served by the RRHs located in the disk of radius $r_1$ centered around its position. We denote this set of RRHs by $\mathcal{C}_{u}$.  All these RRHs are assumed to perfectly estimate their downlink channels before transmitting. The estimated channel values are systematically sent to the BBU. 

\begin{figure}[h!]
    \centering
    \includegraphics[width = 0.34 \textwidth]{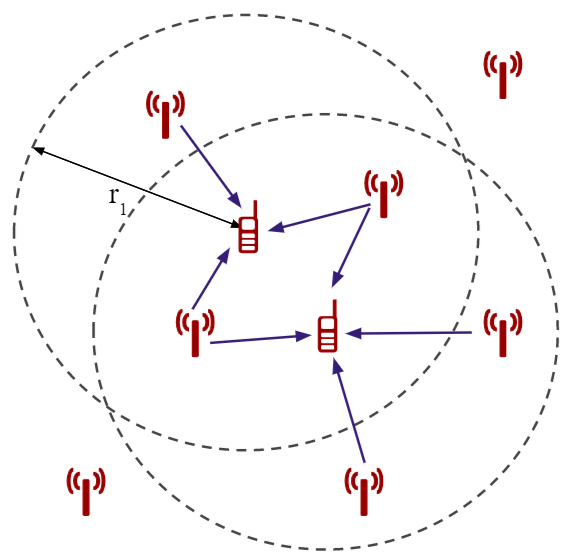}
    \caption{Illustration of the cluster overlaps with distance-based clusters.}
    \label{Association}
\end{figure}

Due to the random node locations, cooperative clusters may overlap (Figure \ref{Association}). A given RRH might thus be associated to several UEs. As mentioned in Section I, such RRH simultaneously transmits a superposition of signals intended to distinct users. In that case, each of these UEs receives both useful information and interference from the same node. In terms of notations, we define $\mathcal{C}_u$ the set of RRHs serving user $u$ and $\mathcal{B}_i$, the set of UEs served by RRH $i$.
\subsection{Downlink signal model}

Without loss of generality, the network performance is evaluated for a centric UE $u^*$ located at $\mathbf{o}=(0,0)$, in accordance with Slivnyak-Mecke theorem \cite{Baccelli}. In the rest of this study, the set $\mathcal{C}_{u^*}$ of RRHs serving this user will be named \textit{centric cluster}. The baseband signal received at that UE is given by
\begin{equation}
    \text{y}_{u^*} = \sum_{i \in \mathcal{C}_{u^*}} \sqrt{P_t} \mathbf{w}_{iu^*}\mathbf{h}_{iu^*} \kappa^{-1/2} r^{-\alpha/2}_{iu^*} s_{u^*} + \sum_{u \in \Psi_{U} \backslash \{u^*\}} \Bigg(\sum_{i \in \mathcal{C}_{u}} \sqrt{P_t} \mathbf{w}_{iu}\mathbf{h}_{iu^*} \kappa^{-1/2} r^{-\alpha/2}_{iu^*} \Bigg) s_u + n_{u^*}
\end{equation}

where the following notations have been introduced:

\begin{itemize}
    \item $P_t$, a constant transmit power;
    \item $\mathbf{w}_{iu} \in \mathbb{C}^{1 \times M}$, the beamforming vector associated to transmission from $i$ to $u$;
    \item $s_u$, the unit energy symbol intended for user $u$;
    \item $n_u$, the additive thermal noise of constant power density $N_0$.
\end{itemize}

Assuming coherent joint transmission, the corresponding received power is given by 
\begin{equation}
\label{eqn_power}
\begin{split}
    P_{u^*} &  = \underbrace{\bigg| \sum_{i \in \mathcal{C}_{u^*}} \sqrt{P_t} \mathbf{w}_{iu^*}\mathbf{h}_{iu^*} \kappa^{-1/2} r^{-\alpha/2}_{iu^*} \bigg|^2}_{P_{S}} + \underbrace{\sum_{u \in \Psi_{U} \backslash \{u^*\}} \bigg| \sum_{i \in \mathcal{C}_{u}} \sqrt{P_t} \mathbf{w}_{iu}\mathbf{h}_{iu^*} \kappa^{-1/2} r^{-\alpha/2}_{iu^*} \bigg|^2}_{P_{I}} + \; N_0
    \end{split} 
\end{equation}
where $P_{S}$ is the useful information power and $P_{I}$ the aggregate interference.

\subsection{Beamforming strategy and power allocation}

For the sake of mathematical tractability, we consider maximum ratio transmission as beamforming strategy. Therefore,
    \begin{equation}
     \label{beamforming_eqn}
        \mathbf{w}_{iu} = \dfrac{\mathbf{g}_{iu}^H}{\gamma_{iu}}
    \end{equation}
    
where $\gamma_{iu}$ is a normalization factor that can be chosen in several manners. In the framework of this paper, this coefficient is defined as $\gamma_{iu} = |\mathbf{g}_{u}|$, with $\mathbf{g}_{u} = [\mathbf{g}_{iu}]_{i \in \mathcal{C}_u}$ being the concatenation of the channel vectors of RRHs serving $u$. This normalization benefits from the following properties:
\begin{itemize}
    \item the effective transmit power allocated to RRH $i$ to serve user $u$ is proportional to the relative channel gain of this RRH with respect to the channel coefficients of all RRHs of $\mathcal{C}_{u}$. This power can indeed be expressed as 
    \begin{equation*}
     \label{Palloc1}
     P^{(u,i)}_{t} = P_t |\mathbf{w}_{iu}s_u|^2 = P_t \dfrac{|\mathbf{g}_{iu}|^2}{|\mathbf{g}_{u}|^2}  = P_t \; \dfrac{|\mathbf{g}_{iu}|^2}{\sum_{i' \in \mathcal{C}_{u}} |\mathbf{g}_{i'u}|^2} 
    \end{equation*}
    \item the total power budget allocated to the RRHs of $C_u$ to serve $u$ is constant and given by
    \begin{equation}
     P^{(u)}_{t} = \sum_{i \in \mathcal{C}_{u}} P^{(u,i)}_{t} = \sum_{i \in \mathcal{C}_{u}} \big| \sqrt{P_t} \mathbf{w}_{iu} s_u \big|^2 = P_t \dfrac{1}{|\mathbf{g}_{u}|^2} \sum_{i \in \mathcal{C}_{u}} |\mathbf{w}_{iu}|^2 = P_t \dfrac{1}{|\mathbf{g}_{u}|^2} \sum_{i \in \mathcal{C}_{u}} |\mathbf{g}_{iu}|^2 = P_t.
\end{equation}
\end{itemize}

From these properties, the power normalization can be interpreted as follows: each UE is granted the same global power $P_t$ by the BBU. This power budget is shared among all RRHs serving this user. Among these RRHs, those experiencing better channel conditions are granted a greater fraction of $P_t$ since their transmitted signals are less attenuated. Other normalization strategies might also be relevant for this scenario and are left for future work. 

\section{Performance Analysis} \label{sect:analysis}

\begin{figure}[h!]
    \centering
    \includegraphics[width=\textwidth]{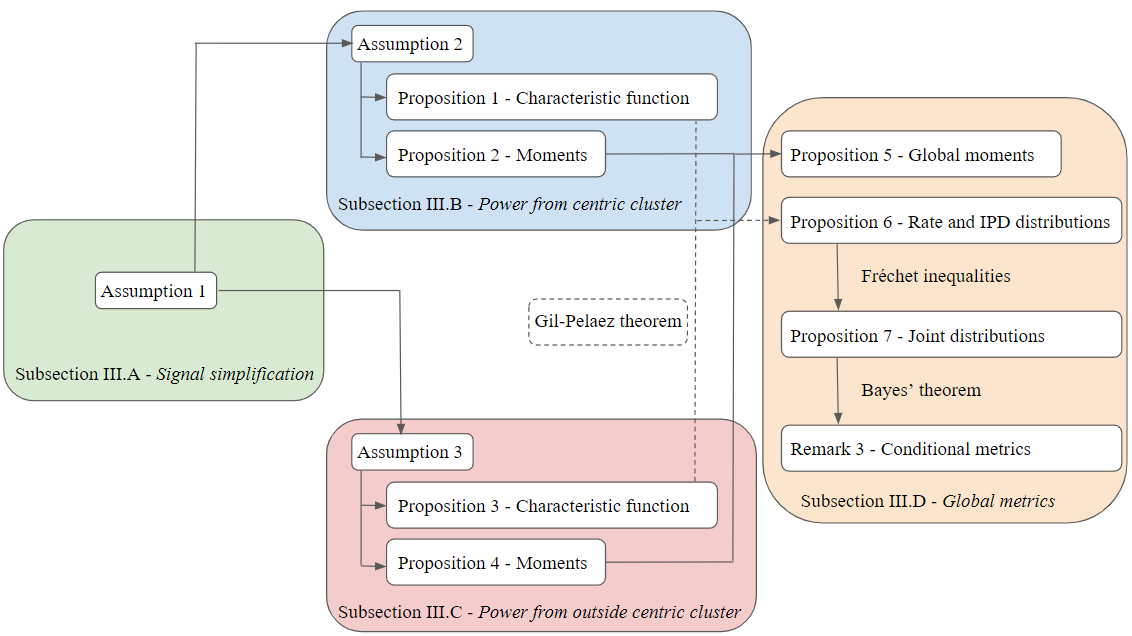}
    \caption{Structure of Section \ref{sect:analysis}.}
    \label{structure_an}
\end{figure}

The structure of this section is illustrated in Figure \ref{structure_an}. A preliminary assumption on the model is provided in Subsection III.A. The incident powers coming from RRHs of the centric cluster and from the other RRHs are studied in Subsections III.B and III.C. Finally, the global performance metrics (main results of this paper) are developed in Subsection III.D.

\subsection{Signal model simplification}

The objective of this first proposition is to simplify the interference term in \eqref{eqn_power} to obtain more adequate expressions for a tractable SG analysis.

\begin{assumption}
\label{CT+moment_matching}
One possible approximation for the aggregate interference $P_I$ is given by 
\begin{equation}
\label{simplified_interference}
        P'_{I} = \underbrace{\sum_{i \in \mathcal{C}_{u^*}} \; \sum_{\substack{u \in \mathcal{B}_i \\ u \neq u^*}} \; P_t  Z^{(i,u)}_{|\mathcal{C}_{u}|} \kappa^{-1} r^{-\alpha}_{iu^*}}_{P_{I_1}} + \underbrace{\sum_{i \in \Psi_{R}\setminus \mathcal{C}_{u^*}} \; \sum_{u \in \mathcal{B}_i} \; P_t Z^{(i,u)}_{|\mathcal{C}_{u}|} \kappa^{-1} r^{-\alpha}_{iu^*}}_{P_{I_2}}.
\end{equation} 

where $\mathcal{B}_i$ denotes the set of UEs served by RRH $i$. The coefficients $Z^{(i,u)}_{|\mathcal{C}_{u}|}$ are random variables following a Gamma distribution of shape and scale parameters given by 

\begin{minipage}{0.49\textwidth}
\begin{align*}
    k_{|\mathcal{C}_{u}|} = \begin{cases}
    1 & \text{if}\; |\mathcal{C}_{u}| = M = 1\\
    \frac{|\mathcal{C}_{u}|M}{3|\mathcal{C}_{u}|-1} & \text{otherwise}
    \end{cases}
\end{align*}

\end{minipage}
\hfill
\hfill
\begin{minipage}{0.49\textwidth}
\begin{align}
    s_{|\mathcal{C}_{u}|} = \begin{cases}
    1 & \text{if}\; |\mathcal{C}_{u}| = M = 1\\
     \frac{3|\mathcal{C}_{u}|-1}{|\mathcal{C}_{u}|^2M} & \text{otherwise}
    \end{cases}
\end{align}
\end{minipage}

\bigskip

\proof
cfr. Appendix \ref{proof_CT+moment_matching}.

\end{assumption}

In (\ref{simplified_interference}), the interference $P'_{I}$ has been separated into two terms: 
\begin{itemize}
    \item $P_{I_1}$, corresponding to the interference power coming from the RRHs of the centric cluster. As explained in section II.C, these RRHs can potentially interfere if they serve other adjacent UEs (cfr. Figure \ref{Association}). This quantity could therefore be described as \textit{intra-cluster interference};
    \item $P_{I_2}$, representing the total interference coming from RRHs located outside the centric cluster. This power could hence be described as \textit{inter-cluster interference.}
\end{itemize}
The objective of the next subsections is to compute the statistics of these two powers. These statistics include their first moments, as well as their characteristic functions (CFs). As illustrated in Figure \ref{structure_an}, these quantities will be employed to derive the global network metrics, including the total IPD moments, as well as the rate and IPD distributions (obtained from the CFs via Gil-Pelaez inversion theorem \cite{GP}).
\subsection{Statistics of the incident power coming from the centric cluster}

\begin{figure}[h!]
     \centering
     \begin{subfigure}[b]{0.33\textwidth}
         \centering
         \includegraphics[width=\textwidth]{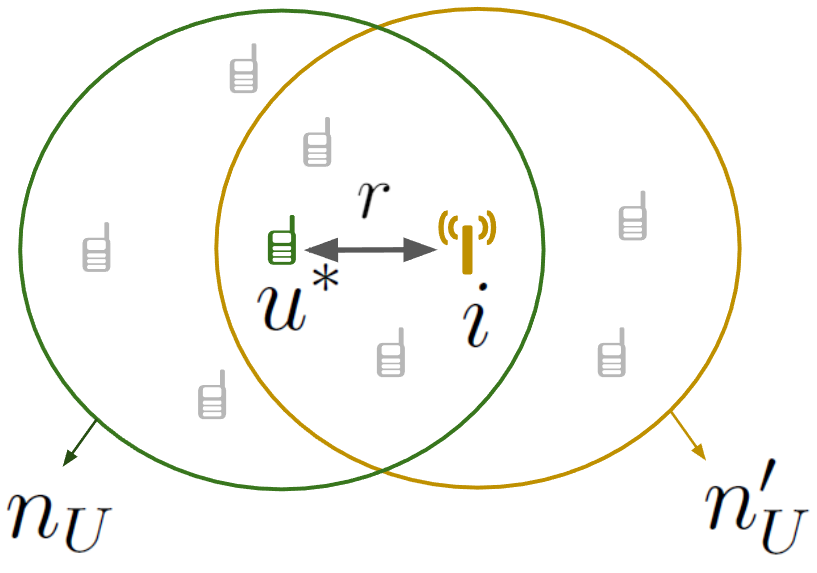}
         \caption{Ilustration of $n_U$ and $n_U'$, corresponding to the number of grey UEs contained in the green and yellow sets.}
         \label{sec_spatial_interf_1}
     \end{subfigure}
     \hspace{1.5 cm}
     \begin{subfigure}[b]{0.4\textwidth}
         \centering
         \includegraphics[width=\textwidth]{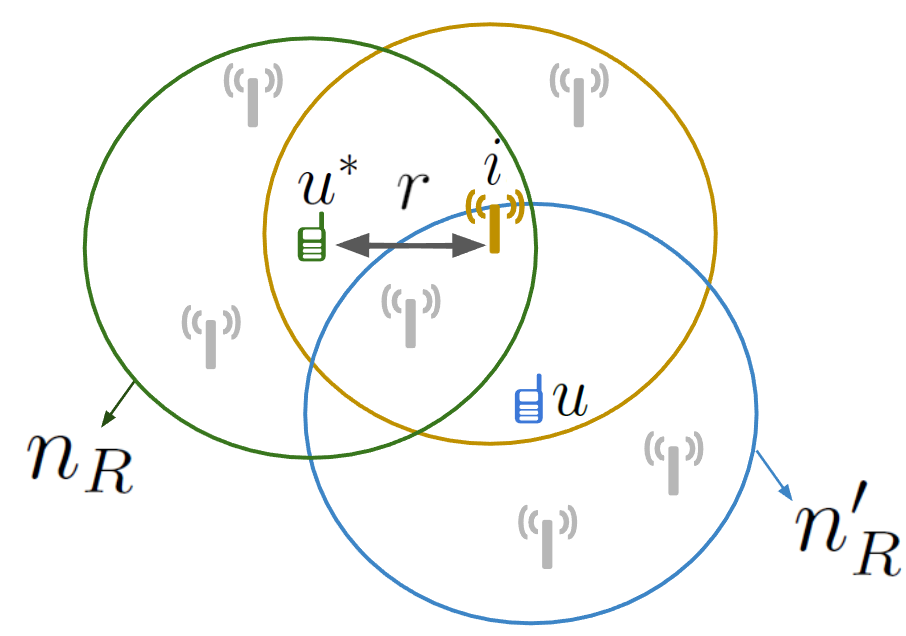}
         \caption{Ilustration of $n_R$ and $n_R'$, corresponding to the number of RRHs contained in the green and blue sets.}
         \label{sec_spatial_interf_2}
     \end{subfigure}
        \caption{Geometrical representation of the quantities $n_U$, $n_U'$, $n_R$ and $n_R'$. All the disk are characterized by a radius of value $r_1$.}
        \label{sec_spatial_interf}
\end{figure}

The interference term $P_{I_1}$ is statistically correlated to the useful received power $P_S$. Both powers share indeed common variables in their definitions, including the number of RRHs in the centric cluster and their locations. Numerical simulations show that deriving the distributions of $P_S$ and $P_{I_1}$ independently leads to a loss of accuracy owing to this correlation (cfr. Subsection IV.A). For this reason, the next proposition provides the characteristic function of the linear combination $P_{S_{\eta}} \triangleq P_S + \eta P_{I_1}$, taking into account the joint statistics of both variables. In order to introduce this proposition, the following elements are defined:

\begin{itemize}
    \item $n_R \triangleq |\mathcal{C}_{u^*}|$, the number of RRHs in the centric cluster (Figure \ref{sec_spatial_interf_2}). This variable follows a Poisson distribution of parameter $\lambda_R\pi(r_1^2-r_0^2)$. The corresponding probability mass function (PMF) is denoted by $p_R(n_R)$;
    
    \item $n_U$, the number of UEs different from $u^*$ located in the disk $\mathcal{D}(\mathbf{o},r_1)$ (Figure \ref{sec_spatial_interf_1}). This random variable also follows a Poisson distribution, of parameter $\lambda_U\pi r_1^2$ and PMF $p_U(n_U)$\footnote[2]{Although $n_R$ and $n_U$ are defined with respect to the centric cluster area, one will observe that their PMFs $p_R(\cdot)$ and $p_U(\cdot)$ can be interpreted with a more global point of view. In all generality, $p_R(\cdot)$ can be defined as the PMF of the number of RRHs serving \textit{any} user in the network. Likewise, for $r_1\gg r_0$, $p_U(\cdot)$ is also the PMF of the number of users served by \textit{any} RRH of $\Psi_R$. In the rest of this paper, these distributions will also be employed with this more general interpretation. }.
\end{itemize}

Considering an arbitrary RRH $i$ of the centric cluster (in yellow in Figure \ref{sec_spatial_interf}), we also define
\begin{itemize}
    \item $n_U' = |\mathcal{B}_i|$, the number of UEs different from $u^*$ served by this RRH (Figure \ref{sec_spatial_interf_1});
    \item $n_R'$, the number of RRHs serving one arbitrary UE $u$ among these $n_U'$ UEs (Figure \ref{sec_spatial_interf_2}).
\end{itemize}
Let us still consider RRH $i$ in Figure \ref{sec_spatial_interf_2}. The interference on $u^*$ coming from this RRH depends on the number of users it serves (variable $n_U'$) and on the effective transmit powers it employs to cover each of these $n_U'$ users. Because of the beamforming policy, these transmit powers are themselves function of the number of RRHs serving each of these $n_U'$ users. In Figure \ref{sec_spatial_interf_2}, this number of RRHs corresponds to $n_R'$ in the case of user $u$ (in blue). Furthermore, one can observe from Figures \ref{sec_spatial_interf_1} and \ref{sec_spatial_interf_2} that $n_U'$ and $n_R'$ are respectively correlated with $n_U$ and $n_R$. To take into account these dependencies, the statistics derived in this section are first conditioned on $n_U$, $n_U'$, $n_R$ and $n_R'$, then progressively averaged: first over $n_U'$ and $n_R'$, and then over $n_U$ and $n_R$. In order to perform such averaging, the next assumption defines conditional probabilities linking these four variables.

\begin{assumption}
\label{pre_corr2}
Conditioned on $n_U$, $n_R$ and distance $r$ (from RRH $i$ to the $u^*$ in Fig. \ref{sec_spatial_interf_2}), the PMFs of $n_U'$ and $n_R'$ are approximated by $p(n_U'|n_U,r)$ and $p(n_R'|n_R,r)$ given in Appendix \ref{proof_pre_corr2}.

\proof
cfr. Appendix \ref{proof_pre_corr2}.
\end{assumption}

\medskip


\begin{proposition}
\label{corr2}
Under the previous assumptions, the characteristic function of $P_{S_{\eta}}$ is given by 
\begin{equation}
\label{SandT}
\phi_{P_{S_{\eta}}}(t) = \sum_{n_U = 0}^{\infty}\sum_{n_R = 0}^{\infty} p_R(n_R)p_U(n_U) \phi_{P_{S_{\eta}}}(t|n_R,n_U) 
\end{equation}

where 
\begin{align}
        \phi_{P_{S_{\eta}}}(t|n_R,n_U) &= \Bigg[\int_{r_0}^{r_1} \phi_{T}(t|n_R,n_U,r)\phi_{V}(\eta t|n_R,n_U,r) \frac{2r}{r_1^2-r_0^2} dr\Bigg]^{n_R} \label{uncondi} \\
        \label{phi_S}
        \phi_{T}(t|n_R,n_U,r) &= \big(1-jtP_t \kappa^{-1}r^{-\alpha}\big)^{-M}\\
        \phi_{V}(t|n_R,n_U,r) &= \sum_{n_U'=0}^{\infty} p(n_U'|n_U,r) \Bigg[\sum_{n_R' = 1}^{\infty} p(n_R'|n_R,r) \; \big(1-jtP_t \kappa^{-1}r^{-\alpha}s_{n_R'}\big)^{-k_{n_R'}} \Bigg]^{n_U'},
        \label{phi_T}
\end{align}
\proof
cfr. Appendix \ref{proof_corr2}.
\end{proposition}

\begin{proposition}
\label{mv_inside}
The mean and the variance of $P_S+P_{I_1}$ are given by
\begin{align}
    &m_{P_S+P_{I_1}}=P_t \kappa^{-1} \sum_{n_U = 0}^{\infty}\sum_{n_R = 0}^{\infty} p_R(n_R)p_U(n_U)n_R L(n_U,n_R)\\
    &\sigma^2_{P_S+P_{I_1}}=P_t^2 \kappa^{-2} \sum_{n_U = 0}^{\infty}\sum_{n_R = 0}^{\infty} p_R(n_R)p_U(n_U) n_R R(n_U,n_R) - m_{P_S+P_{I_1}}^2
\end{align}
where the following auxiliary functions are introduced:
\begin{align*}
    L(n_U,n_R) =& \int_{r_0}^{r_1}\bigg[(M+ \sum_{n_U' = 0}^{\infty} p(n_U'|n_U,r) n_U' \sum_{n_R' = 0}^{\infty} p(n_R'|n_R,r) n_R'^{-1}\bigg] r^{-\alpha} \dfrac{2r}{r_1^2-r_0^2} dr \\
    R(n_U,n_R) =& \int_{r_0}^{r_1} \Bigg[M(M+1) +  2M \sum_{n_U' = 0}^{\infty} p(n_U'|n_U,r) n_U' \sum_{n_R' = 0}^{\infty} p(n_R'|n_R,r) n_R'^{-1} \Bigg] \frac{2r^{-2\alpha+1}}{r_1^2-r_0^2} dr \\
    &+ \int_{r_0}^{r_1} \Bigg[\sum_{n_U' = 0}^{\infty} p(n_U'|n_U,r) n_U' \sum_{n_R' = 0}^{\infty} p(n_R'|n_R,r)  s_{n_R'}^2\dfrac{\Gamma(2+k_{n_R'})}{\Gamma(k_{n_R'})}\Bigg] \frac{2r^{-2\alpha+1}}{r_1^2-r_0^2} dr \\
    &+ \int_{r_0}^{r_1} \Bigg[\sum_{n_U' = 0}^{\infty} p(n_U'|n_U,r) n_U'(n_U'-1) \bigg(\sum_{n_R' = 0}^{\infty} p(n_R'|n_R,r) n_R'^{-1} \bigg)^2\Bigg] \frac{2r^{-2\alpha+1}}{r_1^2-r_0^2} dr  \\
    &+ (n_R - 1) \Bigg( \int_{r_0}^{r_1}\bigg[M+ \sum_{n_U' = 0}^{\infty} p(n_U'|n_U,r) n_U' \sum_{n_R' = 0}^{\infty} p(n_R'|n_R,r) n_R'^{-1}\bigg] \frac{2r^{-\alpha+1}}{r_1^2-r_0^2} dr \Bigg)^2.
\end{align*}

\proof
cfr. Appendix \ref{proof_mean1}.
\end{proposition}

\subsection{Statistics of the power coming from outside the centric cluster}

In order to derive the statistics of $P_{I_2}$, a third assumption is introduced, enabling to perform a spatial decomposition of the interferring RRHs.

\begin{assumption}
Let $\Tilde{\Psi}_R = \Psi_R \setminus \mathcal{C}_{u^*}$ be the set of RRHs generating $P_{I_2}$. This set can be decomposed as
\begin{equation}
    \Tilde{\Psi}_R = \bigcup\limits_{n=0}^{\infty} \Tilde{\Psi}_{R,n}
\end{equation}
where $\Tilde{\Psi}_{R,n}$ is the set of RRHs serving exactly $n$ users. Using the displacement theorem \cite{Baccelli}, each $\Tilde{\Psi}_{R,n}$ can be approximated as a homogeneous PPP of reduced intensity $\lambda_{R,n} =\lambda_{R} \; p_U(n)$ with $p_U(n)$, the probability for an arbitrary RRH to serve $n$ users.
\end{assumption}

\begin{proposition}
\label{outside}
Using the above decomposition, the CF of $P_{I_2}$ can be expressed as 
\begin{equation}
   \phi_{P_{I_2}}(t) = \prod\limits_{n=1}^{\infty} \;\; \exp\Bigg\{-2\pi \lambda_{R,n}\int_{r_1}^{\infty} \bigg[1 - \Big(\sum_{m=1}^{\infty} \Tilde{p}_R(m) (1-jtP_t \kappa^{-1}r^{-\alpha}s_m)^{k_m} \Big)^n \bigg] rdr \Bigg\}
   \label{eq_of_PI2}
\end{equation}
where 
\begin{equation}
\label{truncated}
    \Tilde{p}_R(m) = \dfrac{\Big[\lambda_R\pi(r_1^2-r_0^2)\Big]^{m}}{m!}\dfrac{e^{-\lambda_R\pi(r_1^2-r_0^2)}}{1 - e^{-\lambda_R\pi(r_1^2-r_0^2)}}.
\end{equation}

\proof See Appendix \ref{proof_outside}.

\end{proposition}

\begin{proposition}
\label{mv_outside}
The mean and the variance of $P_{I_2}$ are given by
\begin{align}
    m_{P_{I_2}} = 2\pi^2 &P_t \kappa^{-1} r_1^{(2-\alpha)}(\alpha-2)^{-1} \lambda_{R} \lambda_U (r_1^2-r_0^2) \bigg(\sum_{m=1}^{\infty}p_R(m)m^{-1}\bigg)  \\
    \sigma^2_{P_{I_2}} =  2\pi &P_t^2 \kappa^{-2} r_1^{(2-2\alpha)}(2\alpha-2)^{-1} \lambda_{R} \\ 
    & \times \Bigg[ \pi \lambda_U (r_1^2-r_0^2) \bigg(\sum_{m=1}^{\infty}p_R(m)\rho_m\bigg) + \bigg(\sum_{n=1}^{\infty}p_U(n)n(n-1)\bigg) \bigg(\sum_{m=1}^{\infty}p_R(m)m^{-1}\bigg)\Bigg] \nonumber
\end{align}
with $\rho_m = s_m^2 \Gamma(2+k_m)/\Gamma(k_m)$.

\proof
The proof consist in applying Campbell theorem \cite{Baccelli} as well as the second-order product density of the PPP \cite{Haenggi_book}. 
\end{proposition}
\subsection{Global metrics}

On the basis of the previous results, the total exposure and data rate distribution can now be fully characterized. 

\begin{proposition}
\label{mv_global}
The mean and the variance of the total IPD are given by
\begin{align}
    m_{P_{S}+P_{I_1}+P_{I_2}} &= m_{P_{S}+P_{I_1}} + m_{P_{I_2}} \\
    \sigma^2_{P_{S}+P_{I_1}+P_{I_2}} &= \sigma^2_{P_{S}+P_{I_1}} + \sigma^2_{P_{I_2}}
\end{align}
\proof the proof follows from Propositions \ref{mv_inside} and \ref{mv_outside}, and from the assumed independence between $P_{S}+P_{I_1}$ and $P_{I_2}$.

\end{proposition}

\begin{proposition}
\label{GP_corr}
The distribution of the spectral efficiency is given by
\begin{align}
    \mathcal{P}_r(\theta) &\triangleq \mathbb{P}\Bigg[\log_2\bigg(1+\dfrac{P_S}{P'_I+N_0}\bigg) > \theta\Bigg]  \nonumber \\
    &= \frac{1}{2} + \frac{1}{\pi}\int_{0}^{\infty}\operatorname{Im}\Big\{\phi_{P_{S_{\eta = -\Tilde{\theta}}}}(t|n_R,n_U) \phi_{P_{I_2}}(-\Tilde{\theta} t) e^{-jt\Tilde{\theta} N_0}\Big\}t^{-1}dt 
     \label{coverage_eq}
\end{align}
with $\Tilde{\theta} = 2^{\theta} - 1$. The IPD distribution is obtained by means of a similar expression:
\begin{align}
    \mathcal{P}_e(\theta) &\triangleq \mathbb{P}\Big[P_S + P'_I < \theta' \Big]      
     = \frac{1}{2} - \frac{1}{\pi}\int_{0}^{\infty}\operatorname{Im}\Big\{\phi_{P_{S_{\eta = +1}}}(t|n_R,n_U) \phi_{P_{I_2}}(t) e^{-jt\theta'}\Big\}t^{-1}dt 
     \label{ipd_eq} 
\end{align}

\proof
See Appendix \ref{proof_GP_corr}.
\end{proposition}

\medskip

\begin{remark}
the coverage probability is trivially derived from (\ref{coverage_eq}) by interchanging $\Tilde{\theta}$ and $\theta$. 
\end{remark}

On the basis of the above marginals, bounds on the joint rate-IPD distributions can now be established. These distributions are given by 
\begin{align}
\mathcal{F}(\theta,\theta') = \mathbb{P}\Big[\log_2 \big(1+ P_S/\big[P_I + N_0 \big] \big) > \theta, P_S+P_I > \theta' \Big] \label{EHbound}\\
\mathcal{G}(\theta,\theta') = \mathbb{P}\Big[\log_2 \big(1 + P_S/\big[P_I + N_0 \big] \big)> \theta, P_S+P_I < \theta' \Big]
\label{safetybound}
\end{align}
where \eqref{EHbound} and \eqref{safetybound} can respectively be employed scenarios of WPT and EMF monitoring for safety. 
\begin{proposition} 
\label{prop_joint}
The joint rate-IPD distributions are bounded by the following expressions:
\begin{align}
\label{fbis}
&\max\Big(0,  \mathcal{P}_r(\theta) - \mathcal{P}_e(\theta') \Big) \leq \mathcal{F}(\theta,\theta') \leq \min\Big(\mathcal{P}_r(\theta), 1  - \mathcal{P}_e(\theta') \Big) \\
&\max\Big(0, \mathcal{P}_r(\theta) + \mathcal{P}_e(\theta') -1  \Big) \leq \mathcal{G}(\theta,\theta') \leq \min\Big( \mathcal{P}_r(\theta), \mathcal{P}_e(\theta')  \Big) 
\label{gbis}
\end{align}

\proof
the results come from the application of Fréchet inequalities \cite{frechet}.

\end{proposition}

\begin{remark}
\label{original_vs_sg}
one will note that these bounds are defined with respect to the modified model of interference $P_I'$, employed to derive $\mathcal{P}_r(\theta)$ and $ \mathcal{P}_e(\theta')$. As a reminder, this modified model takes into account assumptions 1 to 3 compared to the original system model of Section II. 
\end{remark}

\begin{remark}
It is also possible to derive bounds for the corresponding conditional distributions. For example, one can define
\begin{equation}
\mathcal{K}(\theta,\theta') =   \mathbb{P}\bigg[ P_S+P_I < \theta' \; \Big| \; \log_2(1+ P_S/P_I) > \theta\bigg].
\end{equation}
In terms of interpretation, this metric represents the probability for a user in a good coverage (spectral efficiency greater than $\theta$) to be affected by a low exposure value (i.e. lower than $\theta'$). Using Bayes rule, bounds for this metric are given by:
\begin{equation}
\dfrac{\max\big(0, \mathcal{P}_r(\theta) + \mathcal{P}_e(\theta') -1  \big)}{ \mathcal{P}_r(\theta)}\leq \mathcal{K}(\theta,\theta') \leq \dfrac{\min\big( \mathcal{P}_r(\theta), \mathcal{P}_e(\theta') \big)}{ \mathcal{P}_r(\theta)}.
\end{equation}

From these distributions, it is possible to derive bounds on conditional means. Building up on the same example, the following conditional mean can be defined:
\begin{equation}
   m_{\theta} = \mathbb{E}\bigg[P_S+P_I \Big| \log_2(1+ P_S/P_I) > \theta\bigg].
\end{equation}
Using the relationship between the mean of a positive random variable and its cumulative distribution function, the following lower and upper bounds on $m_{\theta}$ can be derived: 
\begin{align*}
    \underline{m}_{\theta} &= \int_{0}^{\infty} 1 - \max\Big(0, \mathcal{P}_r(\theta) + \mathcal{P}_e(\theta') -1  \Big)\mathcal{P}_r(\theta)^{-1}  d\theta' \\
    \overline{m}_{\theta} &= \int_{0}^{\infty} 1 - \min\Big( \mathcal{P}_r(\theta), \mathcal{P}_e(\theta') \Big) \mathcal{P}_r(\theta)^{-1}  d\theta'
\end{align*}

This remark is valid for all other conditional distributions and means associated to \eqref{fbis} and \eqref{gbis}. These other combinations are not detailed here due to space limitations.
\end{remark}
\section{Numerical Results}
\label{section_numerical_results}

In this section, the following aspects are successively investigated: influence of the cluster overlaps, impact of the RRH density, rate-IPD trade-offs, and optimal joint performance. For the sake of readability, the densities are here expressed in terms of average numbers of RRHs and UEs per cluster area (denoted by $n_{av,R}$ and $n_{av,U}$), instead of actual values $\lambda_R$ and $\lambda_U$ in $m^{-2}$. The relationships between these quantities are given by $n_{av,R} = \lambda_R \pi (r_1^2-r_0^2)$ and $n_{av,U} = \lambda_U \pi r_1^2$.

\subsection{Impact of the cluster overlaps}
\label{cluster_overlap_impact}
\begin{figure}[h!]
    \centering
    \includegraphics[width = 0.5\textwidth]{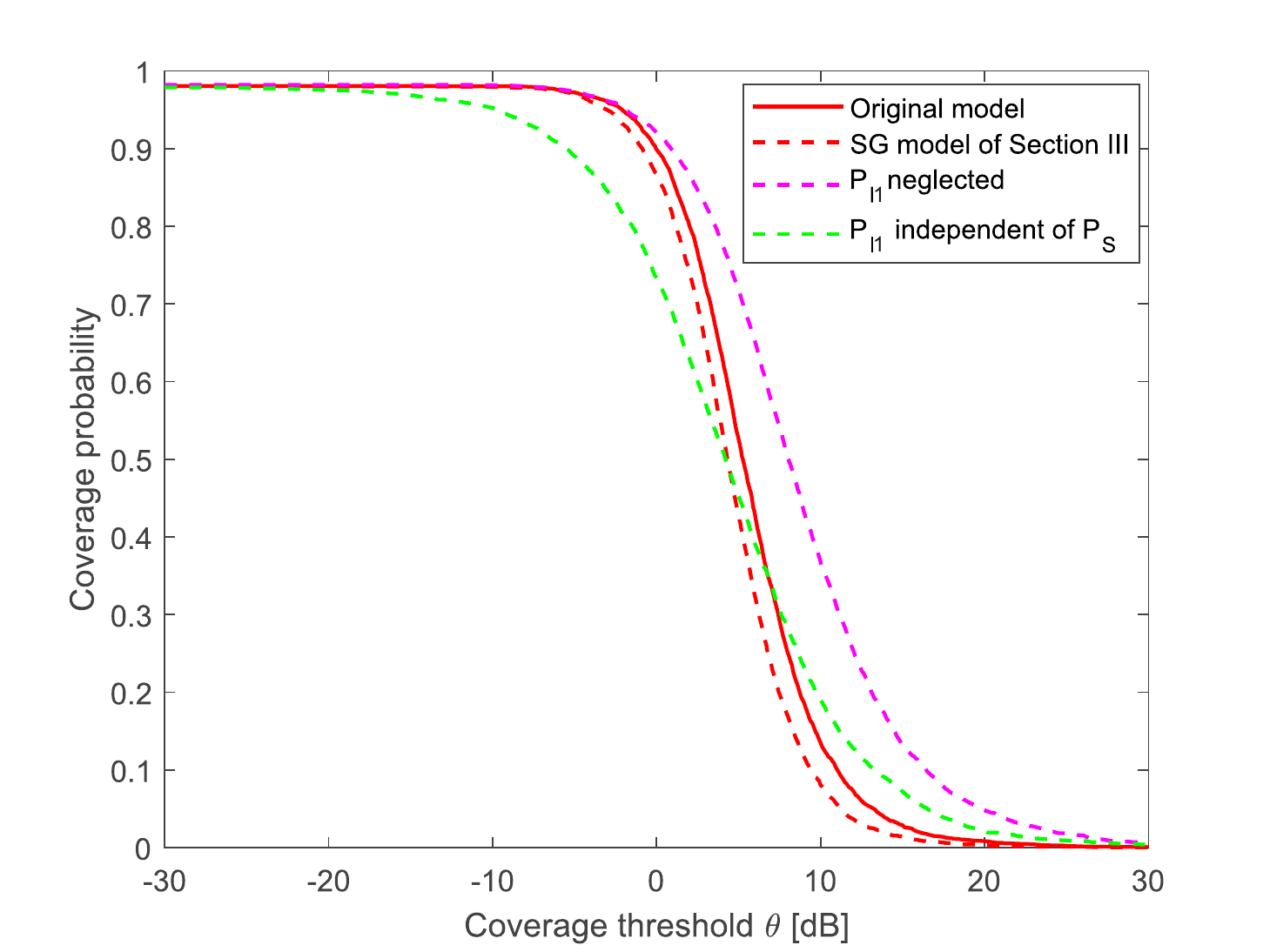}
    \caption{Impact of the cluster overlaps. $P_t = 100 \; mW$, $f = 3\; GHz$, $r_0 = 1\; m$, $r_1 = 100 \; m$, $n_{av,R} = 4$,  $n_{av,U} = 2.5$, $\alpha = 2.5$, $N_0 = 0$ and $M= 4$. }
    \label{cluster_overlap_graph}
\end{figure}

Figure \ref{cluster_overlap_graph} illustrates the coverage probability obtained for the original system model of Section \ref{sect:system}, along with three possible analytical adaptations: 
\begin{itemize}
    \item the full SG approach proposed in this work and detailed in Section \ref{sect:analysis};
    \item the model obtained if the intra-cluster interference (coming from RRHs serving the centric user) is not taken into account. This particular case is obtained from our results by setting $\eta$ to $0$ instead of $-\Tilde{\theta}$ in \eqref{coverage_eq}, which enables to virtually set $P_{I_1}$ to zero;
    \item a third case assuming that the intra-cluster interference is not neglected, but considered as a random variable independent of the useful power $P_S$. One possibility for this scenario can be obtained by modelling the intra-cluster interference in the same manner as the inter-cluster interference $P_{I_2}$. To this purpose, two modifications must be applied to the developments of section \ref{sect:analysis}: $P_{I_1}$ must be set to zero (as in the previous case) and the spatial zone associated to $P_{I_2}$ must be extended to include the disk of radius $r_1$ around the centric user. This second modification is trivially obtained by replacing $r_1$ by $r_0$ in \eqref{eq_of_PI2}.
\end{itemize}
One will note the loss of accuracy obtained when the overlaps are not properly modelled. The curve derived when the intra-cluster interference is neglected naturally leads to higher coverage values compared to the other models. In addition, one can note that treating $P_{I_1}$ independently of $P_S$ produces a distribution centered around the same values as the original model, but with a different variance. Finally, one will note a slight gap remaining between the original system model of section II and the SG results of section III due to Assumptions 1 to 3.
\subsection{Marginal distributions and RRH density}
\label{marginals_num}

\begin{figure}[h!]
     \centering
     \begin{subfigure}[b]{0.49\textwidth}
         \centering
         \includegraphics[width=\textwidth]{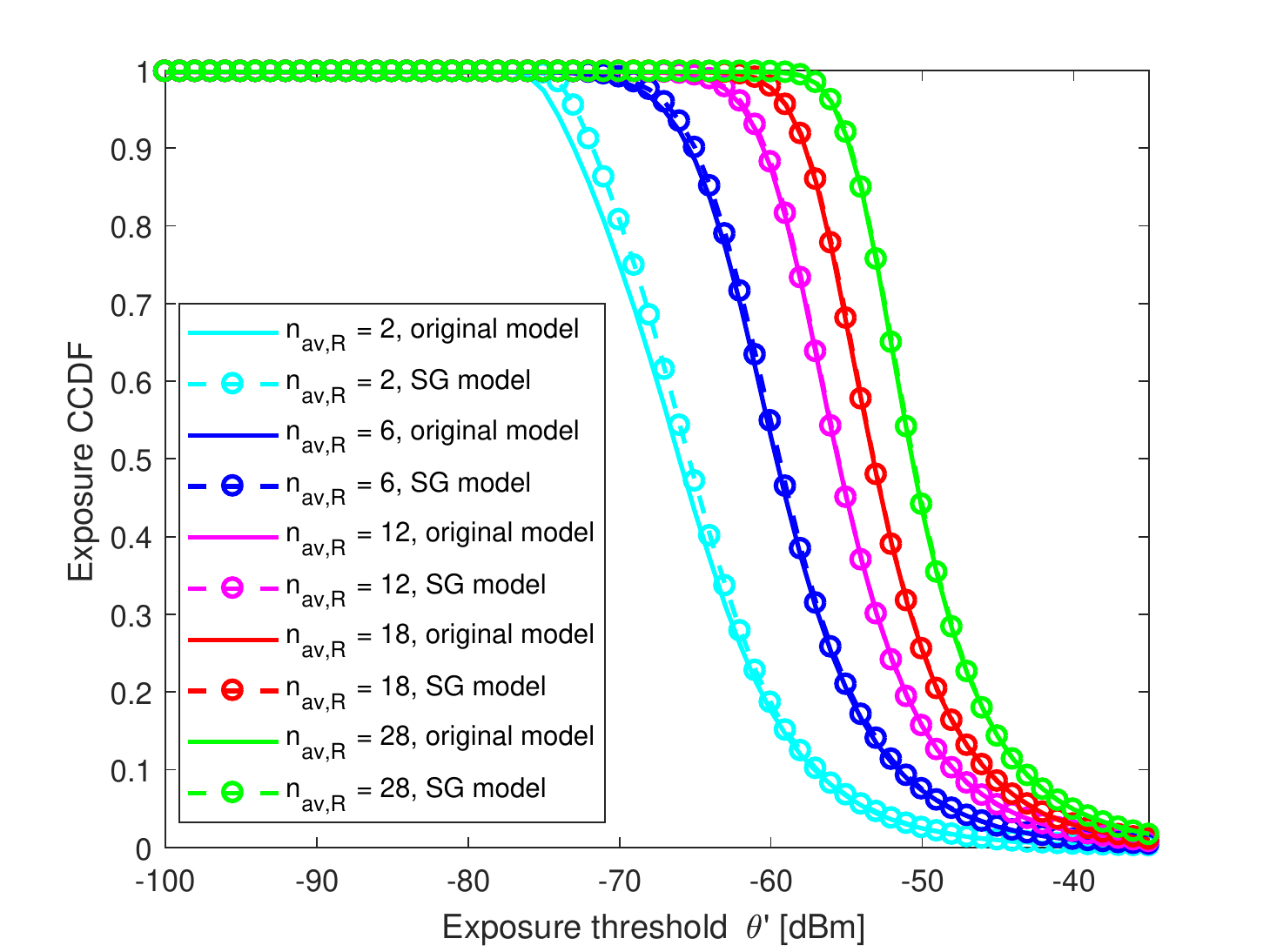}
         \caption{Exposure distribution.}
     \end{subfigure}
     \hfill
     \begin{subfigure}[b]{0.49\textwidth}
         \centering
         \includegraphics[width=\textwidth]{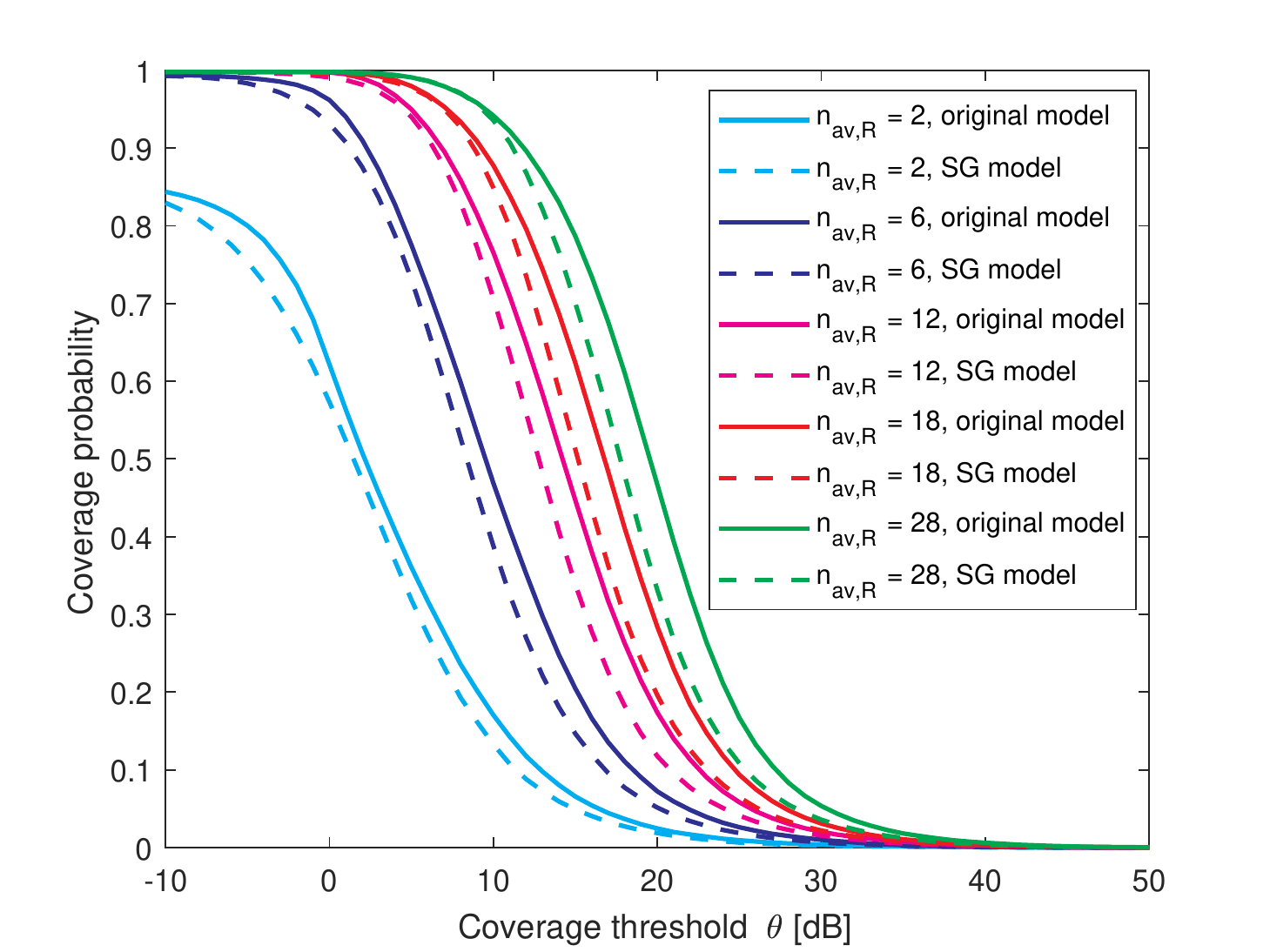}
         \caption{Coverage probability.}
     \end{subfigure}
        \caption{Coverage and exposure cumulative distributions for several RRH densities, and for $P_t = 100 \; mW$, $f = 4 \; GHz$, $r_0 = 1\; m$, $r_1 = 100 \; m$, $n_{av,U} = 0.5$, $\alpha = 2.5$, $N_0 = 0$ and $M= 1$.}
        \label{marginals}
\end{figure}

In Figure \ref{marginals}, the cumulative distributions of the IPD and coverage are displayed for different RRH densities. Several elements can be observed from these graphs: 
\begin{itemize}
    \item The coverage and IPD tend to take higher values as the number of serving RRHs increases. These tendencies can be explained by the power allocation: as a reminder, each user is granted a constant power budget distributed among its serving RRHs. Among these RRHs, more power is allocated to those in good propagation conditions. Increasing the number of serving nodes results in obtaining candidates closer and closer to the user. These candidates are likely to benefit from better channel conditions (with a lower path loss), and are therefore granted a larger share of the power budget. This increases the useful received power, and subsequently the SINR and exposure values.
    \item This increase tends to slow for higher RRH numbers (e.g. 18 and 28). In terms of interpretation, one can conclude that as the number of RRHs increases, a given improvement in SINR requires more and more additional nodes to obtain new candidates in better conditions.
\end{itemize}

\subsection{Rate-IPD trade-offs}
\label{tdsection}
\begin{figure}[h!]
     \centering
     \begin{subfigure}[b]{0.49\textwidth}
         \centering
         \includegraphics[width=\textwidth]{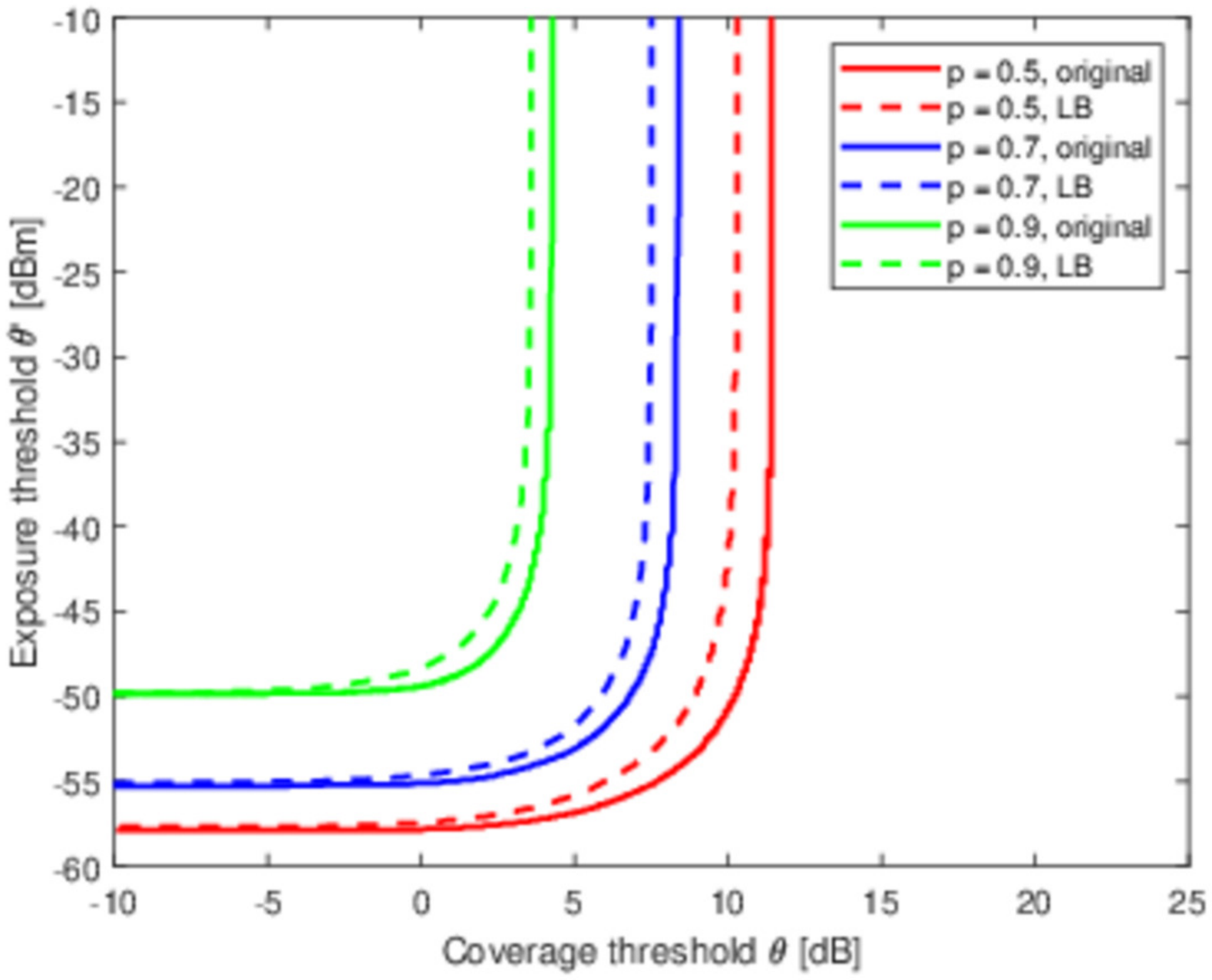}
         \caption{Isocurves of the distribution $\mathcal{G}(\theta,\theta')$ shown for different isovalues $p$. Both the curves associated to the original system model and the lower bound are shown.}
         \label{iso}
     \end{subfigure}
     \hfill
     \begin{subfigure}[b]{0.49\textwidth}
         \centering
         \includegraphics[width=\textwidth]{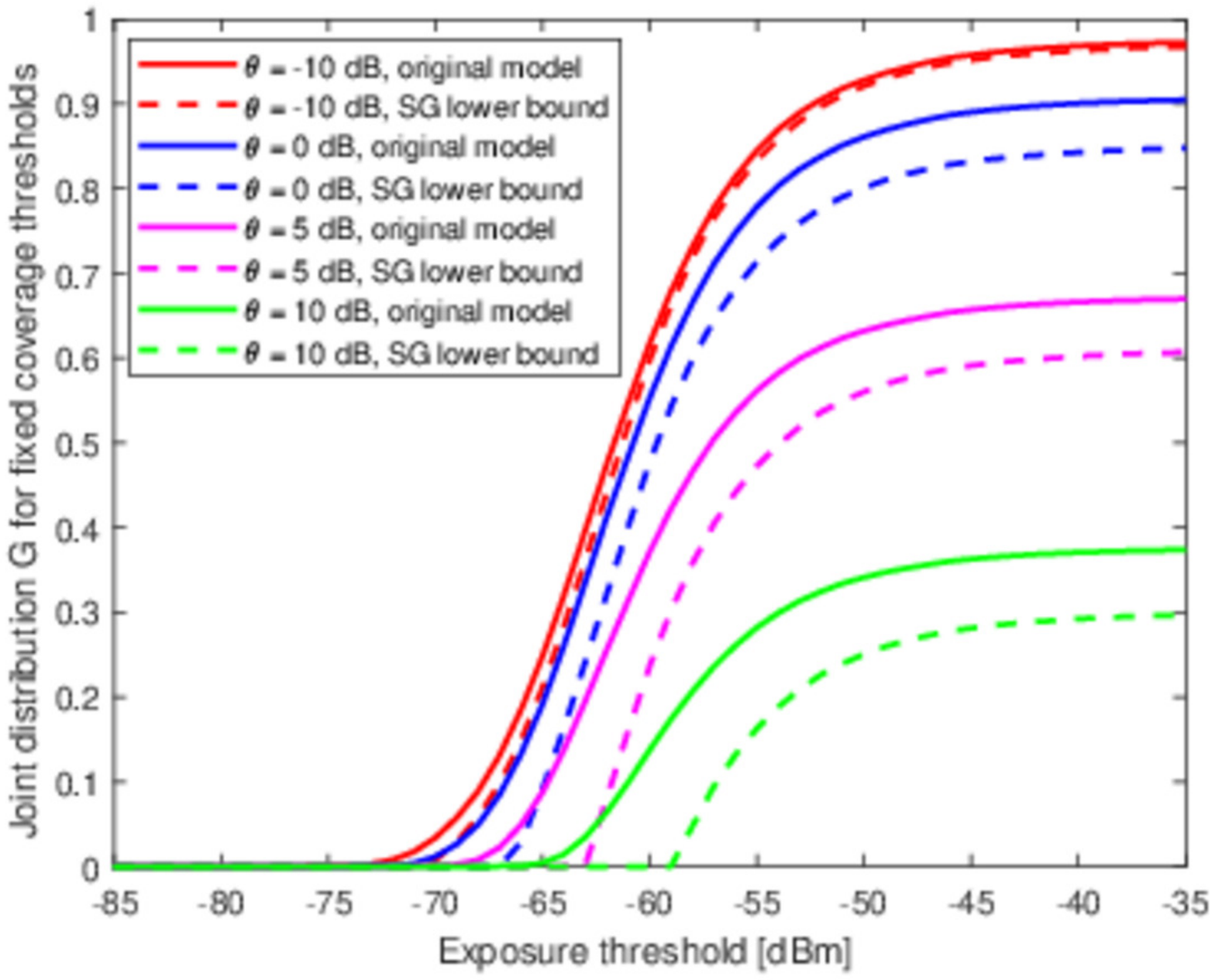}
         \caption{Joint distribution $\mathcal{G}(\theta,\theta')$ as function of the exposure threshold $\theta'$, for fixed SIR thresholds ranging from $-10$ to $10$ dB.}
         \label{result_5}
     \end{subfigure}
        \caption{Rate-IPD trade-offs. Parameters: $P_t = 100 \; mW$, $f = 4 \; GHz$, $r_0 = 1\; m$, $r_1 = 100 \; m$, $n_{av,U} = 0.5$, $\alpha = 2.5$, $N_0 = 0$ and $M= 1$ $n_{av,R} = 8$ for (a) and $n_{av,R} = 4$ for (b).}
        \label{rrrrr}
\end{figure}

Figure \ref{iso} represents isocurves observable when computing the joint metric $\mathcal{G}(\theta,\theta')$. These curves illustrate the trade-offs limiting the coverage and IPD performance. The existence of such compromise is easily explained by considering the useful power $P_S$, which is likely to be high for users experiencing a good coverage. Since this useful power is generated by the RRHs which are the closest to the user, it is also dominant in the total exposure in a large number of cases. The probability of simultaneously experiencing a low exposure and a good coverage therefore decreases as the requirements on $(\theta,\theta')$ become more demanding. The same explanation also justifies the behaviour of $\mathcal{G}(\theta,\theta')$ for fixed coverage thresholds (Figure \ref{result_5}).

\subsection{Existence of an optimal node density}

\begin{figure}[h!]
     \centering
     \begin{subfigure}[b]{0.32\textwidth}
         \centering
         \includegraphics[width=\textwidth]{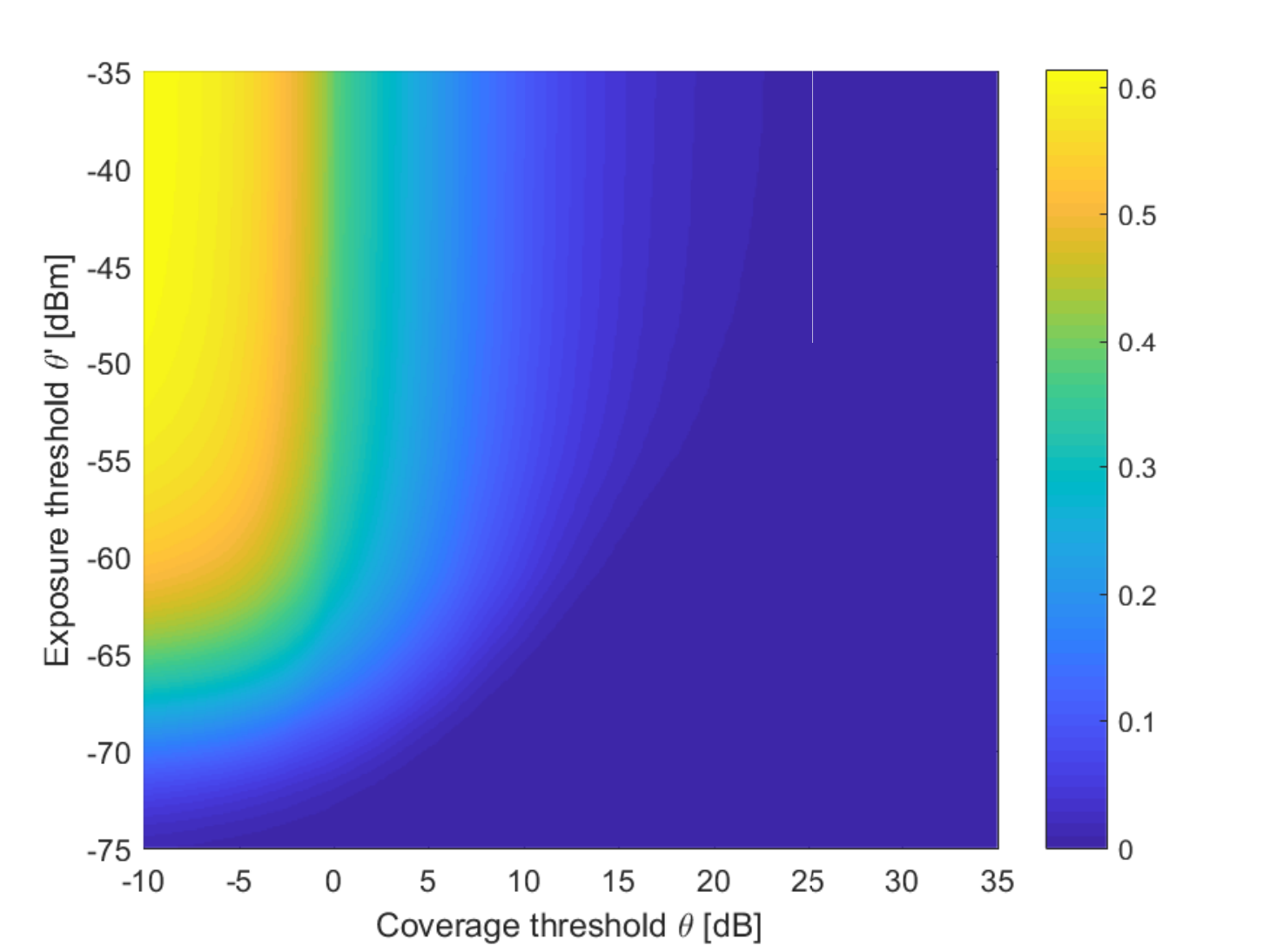}
         \caption{$n_{av,R} = 1$}
         \label{dem1}
     \end{subfigure}
     \hfill
     \begin{subfigure}[b]{0.32\textwidth}
         \centering
         \includegraphics[width=\textwidth]{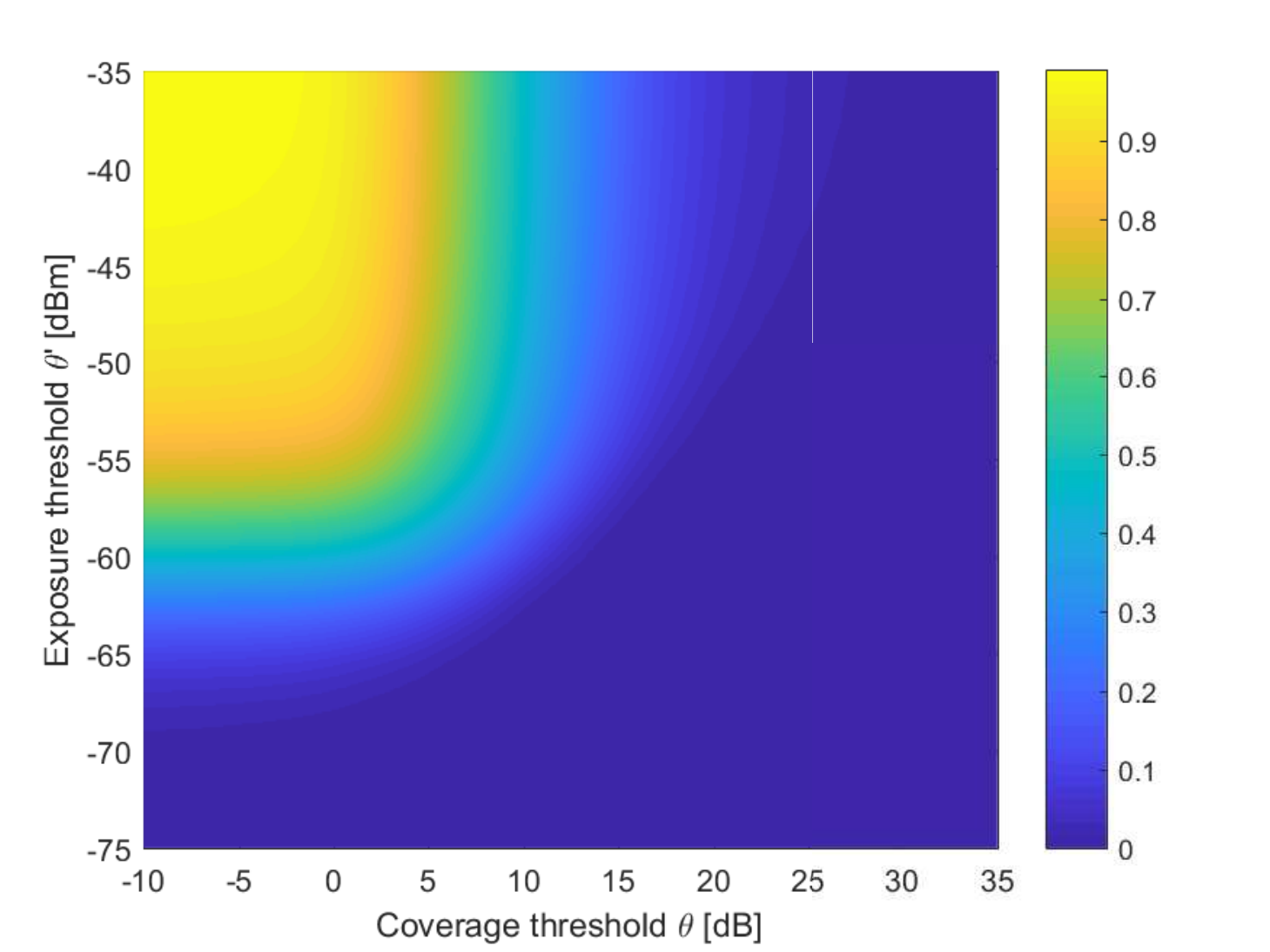}
         \caption{$n_{av,R} = 6$}
         \label{dem2}
     \end{subfigure}
     \begin{subfigure}[b]{0.32\textwidth}
         \centering
         \includegraphics[width=\textwidth]{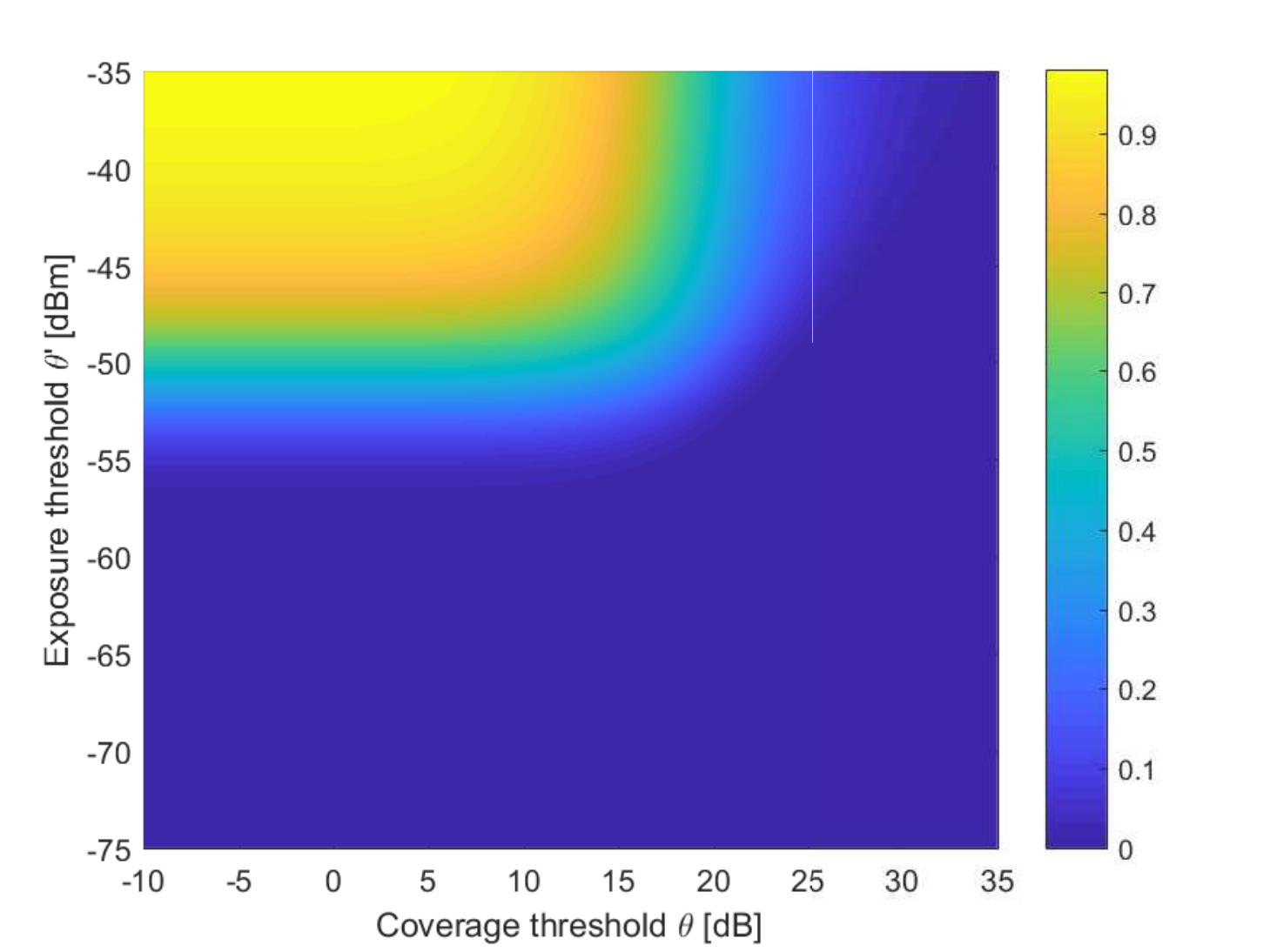}
         \caption{$n_{av,R} = 30$}
         \label{dem3}
     \end{subfigure}
        \caption{Evolution of the metric $\mathcal{G}(\theta,\theta')$ as function of the average number of RRHs per cluster. The system parameters are here given by $P_t = 100 \; mW$, $f = 4 \; GHz$, $r_0 = 1\; m$, $r_1 = 100 \; m$, $n_{av,U} = 0.5$, $\alpha = 2.5$, $N_0 = 0$ and $M= 1$.}
        \label{dem}
\end{figure}

\begin{figure}
     \centering
     \begin{subfigure}[b]{0.45\textwidth}
         \centering
         \includegraphics[width=\textwidth]{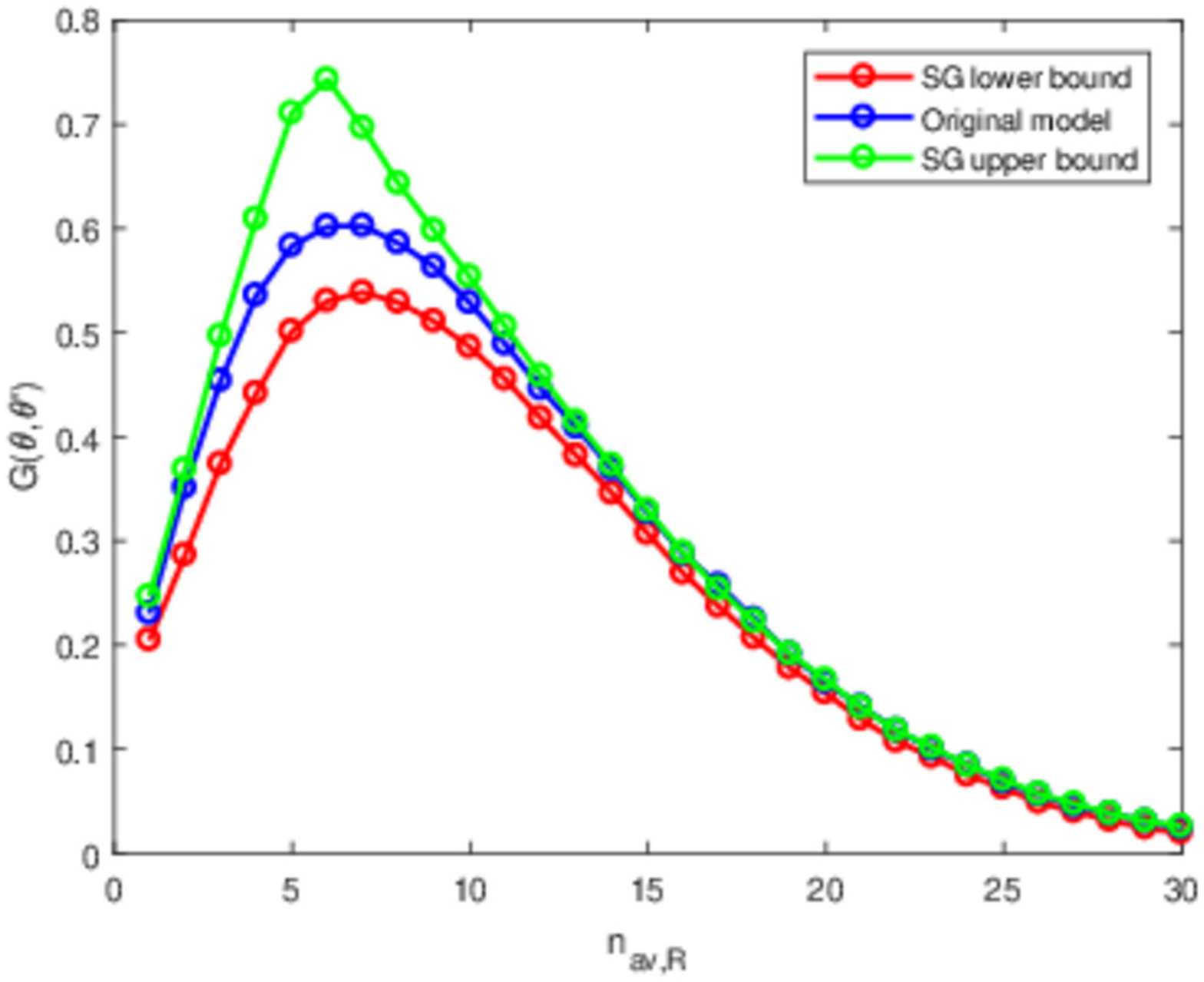}
         \caption{$(\theta,\theta') = (5 \; dB, -55\; dBm) $}
         \label{optimals_1}
     \end{subfigure}
     \hfill
     \begin{subfigure}[b]{0.45\textwidth}
         \centering
         \includegraphics[width=\textwidth]{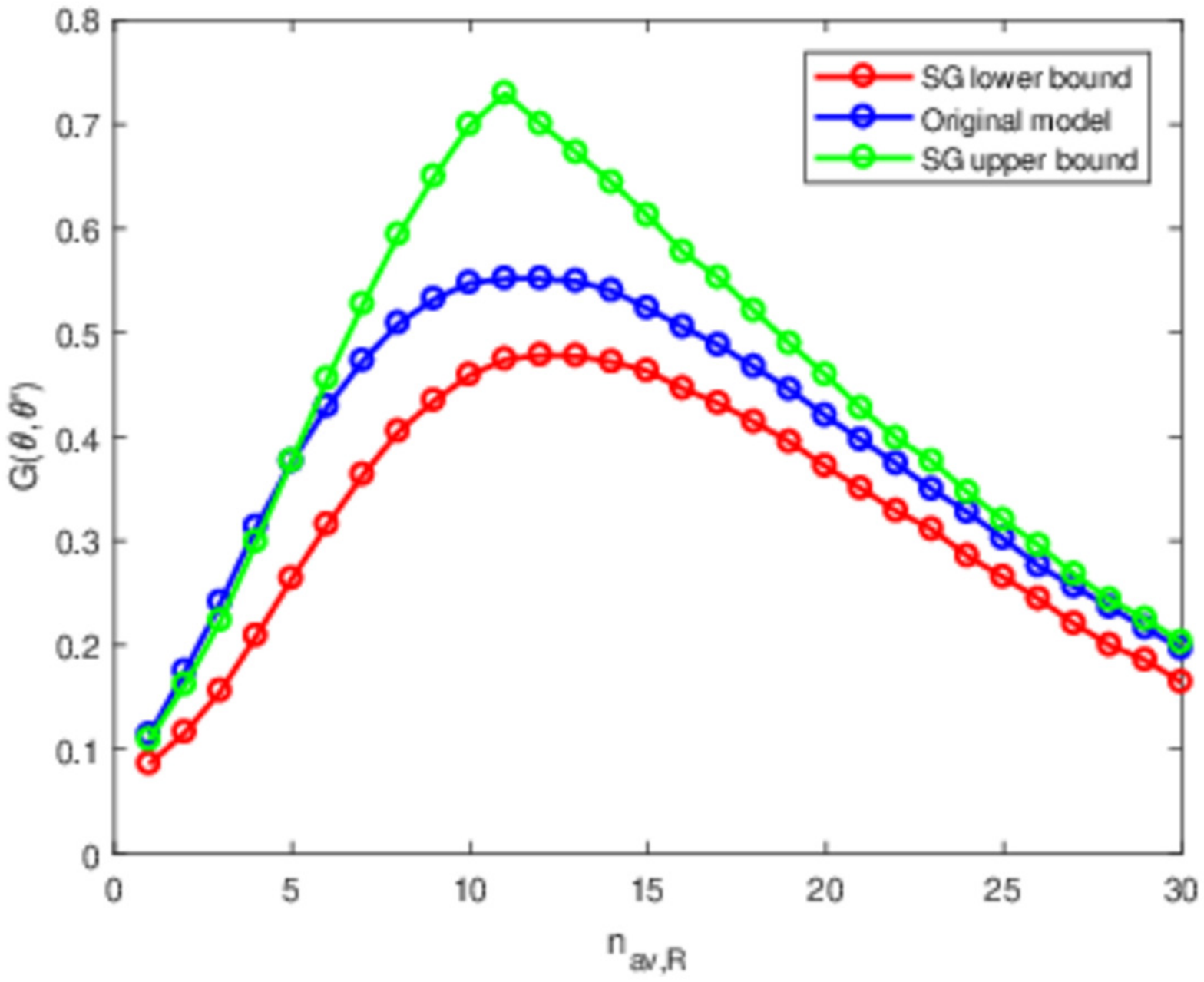}
         \caption{$(\theta,\theta') = (10 \; dB, -52\; dBm) $}
         \label{optimals_2}
     \end{subfigure}
        \caption{Illustration of the optimal RRH densities maximizing $\mathcal{G}(\theta,\theta')$. The system parameters are here given by $P_t = 100 \; mW$, $f = 4 \; GHz$, $r_0 = 1\; m$, $r_1 = 100 \; m$, $n_{av,U} = 0.5$, $\alpha = 2.5$, $N_0 = 0$ and $M= 1$.}
        \label{optimals}
\end{figure}

Figures \ref{dem} depicts $\mathcal{G}(\theta,\theta')$ as function of both $\theta$ and $\theta'$, for increasing RRH densities. For a low density (Figure \ref{dem1}), high probability values are located in the left zone of the graph, characterized by low coverage values. For increasing RRH densities (Figures \ref{dem2} and \ref{dem3}), this yellow zone of high probabilities evolves to eventually lie in an area characterized by a wider interval of coverage thresholds reachable, but with a narrower interval of exposure requirements satisfied. This tendency can be explained using the arguments mentioned in Subsection \ref{tdsection}. In addition, this evolution suggests the existence of an optimal RRH density, different for every thresholds combination $(\theta,\theta')$. Figure \ref{optimals} confirms this hypothesis with two examples of threshold requirements. 

\begin{remark}
As mentioned in Remark \ref{original_vs_sg}, the lower and upper bounds are computed from the SG model relying on Assumptions 1 to 3. Owing to these assumptions, the upper bound in Figure \ref{optimals_2} goes slightly below the curve of original model for low densities. Despite this observation, the two bounds enable to find optimal densities that are relatively accurate compared to the original model.
\end{remark}

\begin{remark}
for sake of space limitations, the analysis of this subsection has been focused on metric $\mathcal{G}(\theta,\theta')$ for which an optimal RRH was highlighted. In a dual manner, this discussion could be transposed to the other metric $\mathcal{F}(\theta,\theta')$, in which case an optimal UE density can be studied. The existence of such optimum can be explained by considering the aggregate interference: bringing additional users in the network has the effect of increasing this interference, which simultaneously decreases the coverage, but increases the total IPD. From the perspective of a network operator, these optima can provide indications on how many RRHs to deploy (resp. users to schedule in a time slot) to maximize the number of UEs satisfying the threshold requirements related to $\mathcal{G}(\theta,\theta')$ (resp. $\mathcal{F}(\theta,\theta')$).
\end{remark}

\section{Conclusion}
This study proposes a comprehensive framework to jointly analyze the rate and IPD in UCCFNs. The complete statistics of these quantities have been derived: moments, marginal, joint and conditional distributions. The provided expressions take into account the important features in the analysis of UCCFNs (cluster intercorrelation, beamforming strategy, etc), some of them being new in SG models.  \\

Possible extensions could include non-uniform node deployments. The homogeneous PPPs considered in this work involve constant UE and RRH densities. Employing an inhomogeneous point process instead could enable to model locations with a more intense activity (hotspots, metro stations,...). Regarding the beamforming, other precoding strategies could be studied and compared with this work. Finally, the incorporation of more advanced channel models could be investigated as well.

\appendices

\section{Proof of Assumption \ref{CT+moment_matching}}
\label{proof_CT+moment_matching}

By injecting (\ref{beamforming_eqn}) in (\ref{eqn_power}), and developing the squared norm, one obtains 
\begin{align}
\label{proof1}
        P_{I} &= P_t \kappa^{-1} \sum_{u \in \Psi_{U} \backslash \{u^*\}} \frac{1}{|\mathbf{g}_{u}|^2}  \sum_{i \in \mathcal{C}_{u}}  \bigg| \mathbf{h}_{iu}^H \mathbf{h}_{iu^*} r^{-\alpha/2}_{iu}  r^{-\alpha/2}_{iu^*} \bigg|^2 \\
        &+ P_t \kappa^{-1} \sum_{u \in \Psi_{U} \backslash \{u^*\}} \frac{1}{|\mathbf{g}_{u}|^2} \sum_{i \in \mathcal{C}_u} \sum_{i' \in \mathcal{C}_{u}, i' \neq i} \underbrace { \mathbf{h}_{iu}^H \mathbf{h}_{iu^*} r^{-\alpha/2}_{iu}  r^{-\alpha/2}_{iu^*} \mathbf{h}_{i'u^*}^H \mathbf{h}_{i'u} r^{-\alpha/2}_{i'u}  r^{-\alpha/2}_{i'u^*} }_{\text{cross terms}}. \label{ct1}
\end{align}

Since the fading is Rayleigh, all the cross terms in the above expression are zero mean. As a result, these terms only impact higher order moments of the aggregate interference. Numerical simulations (see Section \ref{section_numerical_results}) show that neglecting these terms has a low impact on the accuracy of the results. These terms are therefore neglected, which enables to obtain a simpler expression:
\begin{align}
        P_{I} &\approx P_t \kappa^{-1} \sum_{u \in \Psi_{U} \backslash \{u^*\}} \frac{1}{|\mathbf{g}_{u}|^2}  \sum_{i \in \mathcal{C}_{u}}  \bigg| \mathbf{h}_{iu}^H \mathbf{h}_{iu^*} r^{-\alpha/2}_{iu}  r^{-\alpha/2}_{iu^*} \bigg|^2 \\
        &\approx P_t \kappa^{-1} \sum_{u \in \Psi_{U} \backslash \{u^*\}} \frac{1}{|\mathbf{g}_{u}|^2}  \sum_{i \in \mathcal{C}_{u}}  \bigg| \Big(\sum_{k=1}^{M} h_{iuk}^* h_{iu^*k}  \Big) r^{-\alpha/2}_{iu}  r^{-\alpha/2}_{iu^*} \bigg|^2. \label{ctagain}
\end{align}

By neglecting again the cross-terms coming from the norm of \eqref{ctagain} and developing the normalization factor $|\mathbf{g}_{u}|^2$, the expression can be further simplified as
\begin{align}
        \label{interm_quotient}
        P_{I} &\approx P_t \kappa^{-1} \sum_{u \in \Psi_{U} \backslash \{u^*\}} \; \sum_{i \in \mathcal{C}_{u}}   \;  \dfrac{\sum_{k=1}^{M} \big| h_{iuk}\big|^2 \big| h_{iu^*k} |^2 r^{-\alpha}_{iu}}{\sum_{i' \in \mathcal{C}_{u}} \sum_{k=1}^{M} \big| h_{i'uk}\big|^2 r^{-\alpha}_{i'u}} \;  r^{-\alpha}_{iu^*}.
\end{align}
In the quotient in (\ref{interm_quotient}), the distances $r_{iu}$ and $r_{i'u}$ represent the distances from two RRHs $i$ and $i'$ serving a UE $u \neq u^*$. A geometrical representation is proposed in Figure \ref{hyp_spatial_interf_1}. By approximating these distances as being equal for $u \neq u^*$ (cfr. Figure \ref{hyp_spatial_interf_2}), one obtains:
\begin{align}
        \label{distances_equal}
        P_{I} &\approx P_t \kappa^{-1} \sum_{u \in \Psi_{U} \backslash \{u^*\}} \; \sum_{i \in \mathcal{C}_{u}}  \; \underbrace{\dfrac{\sum_{k=1}^{M} \big| h_{iuk}\big|^2 \big| h_{iu^*k} |^2}{\sum_{i' \in \mathcal{C}_{u}} \sum_{k=1}^{M} \big| h_{i'uk}\big|^2}}_{Z^{(i,u)}_{|\mathcal{C}_{u}|}} \;   r^{-\alpha}_{iu^*} 
\end{align}
\begin{figure}
     \centering
     \begin{subfigure}[b]{0.49\textwidth}
         \centering
         \includegraphics[width=\textwidth]{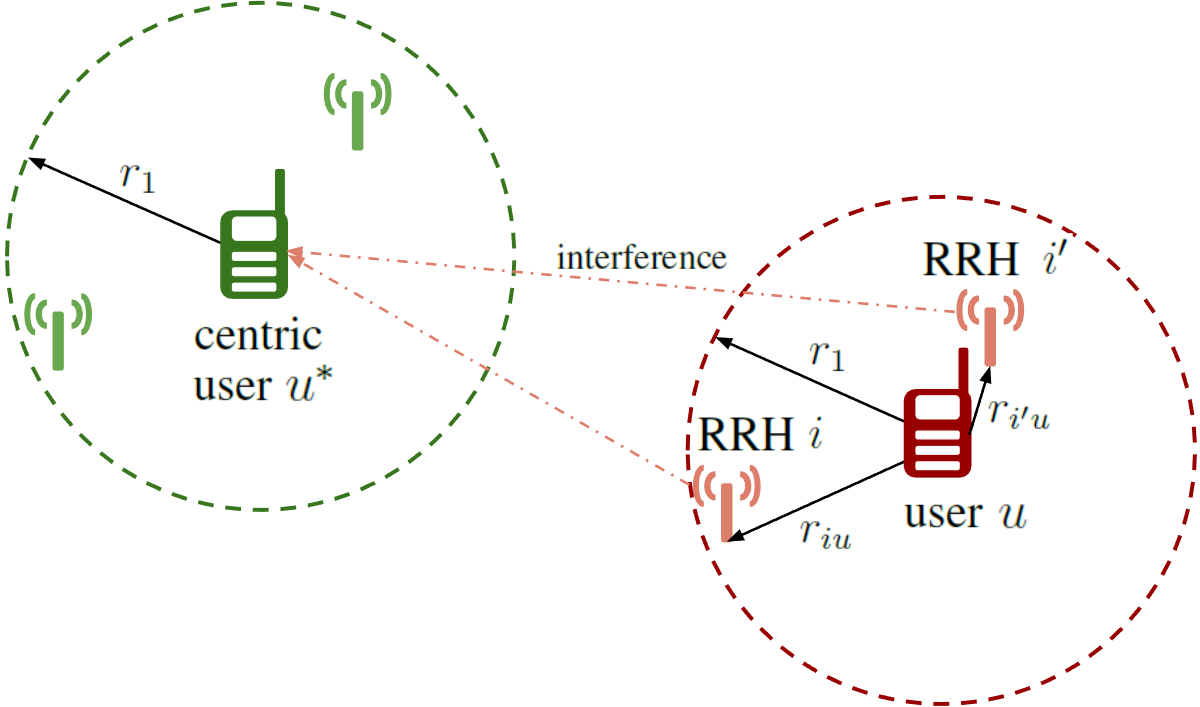}
         \caption{Original topology with $r_{iu} \neq r_{i'u}$.}
         \label{hyp_spatial_interf_1}
     \end{subfigure}
     \hfill
     \begin{subfigure}[b]{0.49\textwidth}
         \centering
         \includegraphics[width=\textwidth]{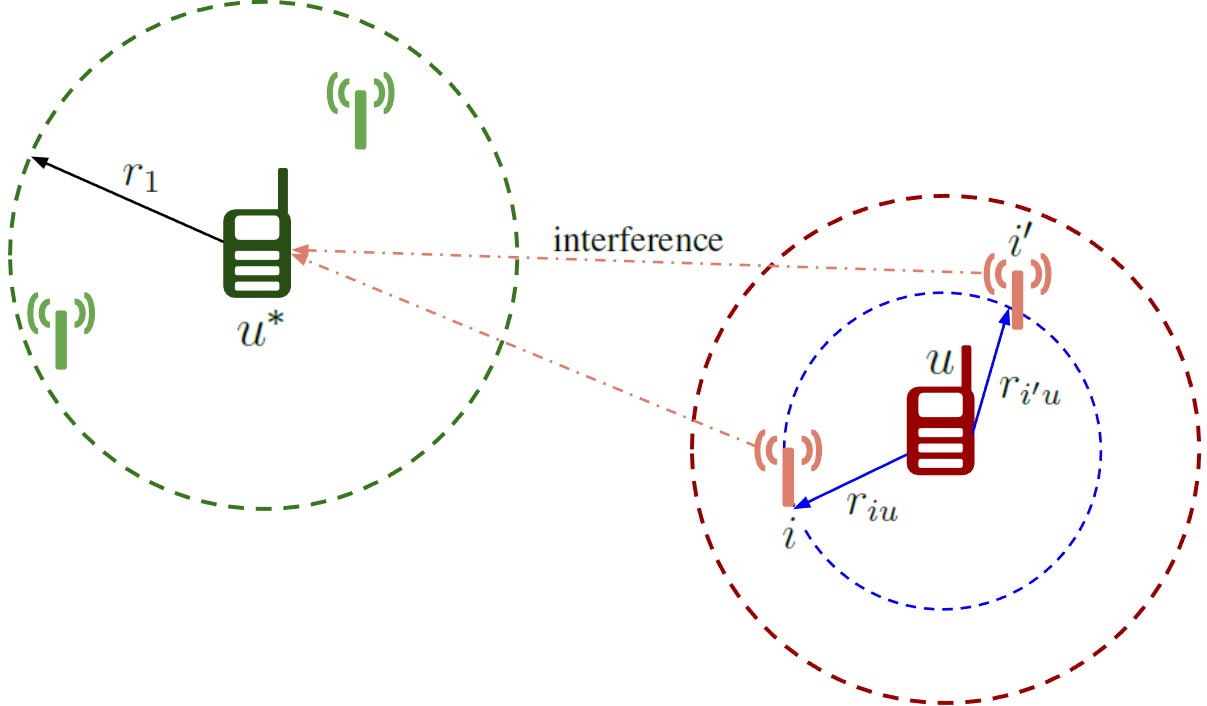}
         \caption{Modified topology with assumption $r_{iu} = r_{i'u}$.}
         \label{hyp_spatial_interf_2}
     \end{subfigure}
        \caption{Illustration of the approximation employed to obtain \eqref{distances_equal}.}
        \label{hyp_spatial_interf}
\end{figure}
The distribution of $Z^{(i,u)}_{|\mathcal{C}_{u}|}$ is then approximated by a Gamma distribution via moment matching. In order to compute the shape and scale parameters of this distribution, the moments of the quotient $Z^{(i,u)}_{|\mathcal{C}_{u}|}$ are approximated using the Taylor expansions presented in \cite{ratio_1,ratio_2,ratio_3}. Equation \eqref{simplified_interference} is finally obtained by rewriting the double summation as functions of sets $\Psi_R$ and $\mathcal{B}_i$.

\section{Proof of Assumption \ref{pre_corr2}}
\label{proof_pre_corr2}

The probabilities $p(n_U'|n_U,r)$ and $p(n_R'|n_R)$ are defined as follows: 
\begin{itemize}
    \item $p(n_U'|n_U,r)$ is the probability for a RRH $i$ belonging to the centric cluster to serve $n_U'$ UEs different from $u^*$, given that $n_U$ UEs different from $u^*$ are in $\mathcal{D}(\mathbf{o},r_1)$ (cfr. Figure \ref{hyp_spatial_interf_1}).
    \item Considering a UE $u$ (different from $u^*$) served by a RRH $i$ of the centric cluster, $p(n_R'|n_R,r)$ represents the probability for this UE to be served by $n_R'$ RRHs, given $n_R$ RRHs in the centric cluster (cfr. Figure \ref{hyp_spatial_interf_2}).
\end{itemize}
\begin{figure}[h!]
     \centering
     \begin{subfigure}[b]{0.38\textwidth}
         \centering
         \includegraphics[width=\textwidth]{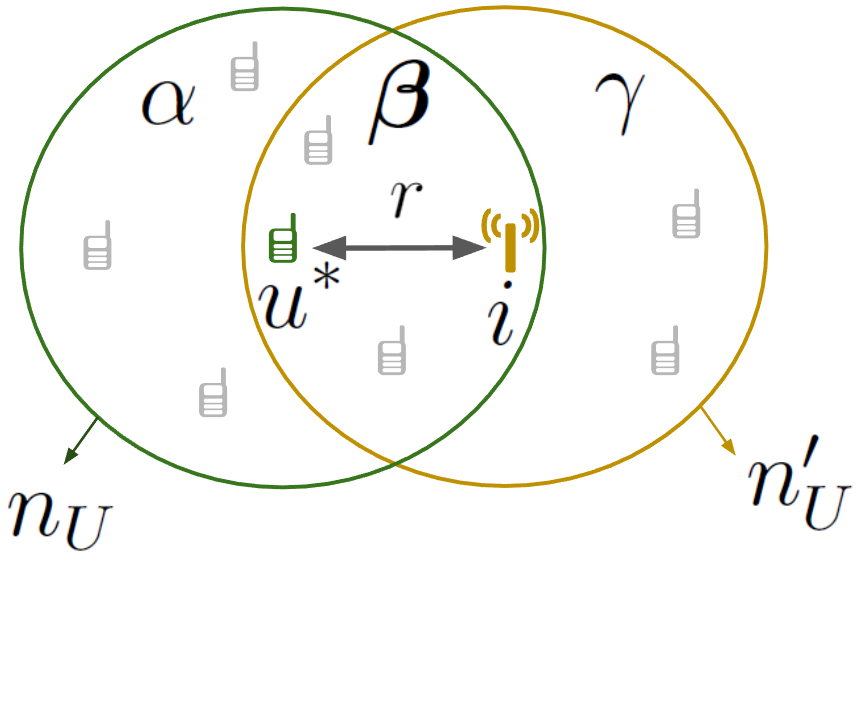}
         \caption{Intersection considered for $p(n_U'|n_U,r)$.}
         \label{hyp_spatial_interf_1}
     \end{subfigure}
     \hspace{0.7 cm}
     \begin{subfigure}[b]{0.4\textwidth}
         \centering
         \includegraphics[width=\textwidth]{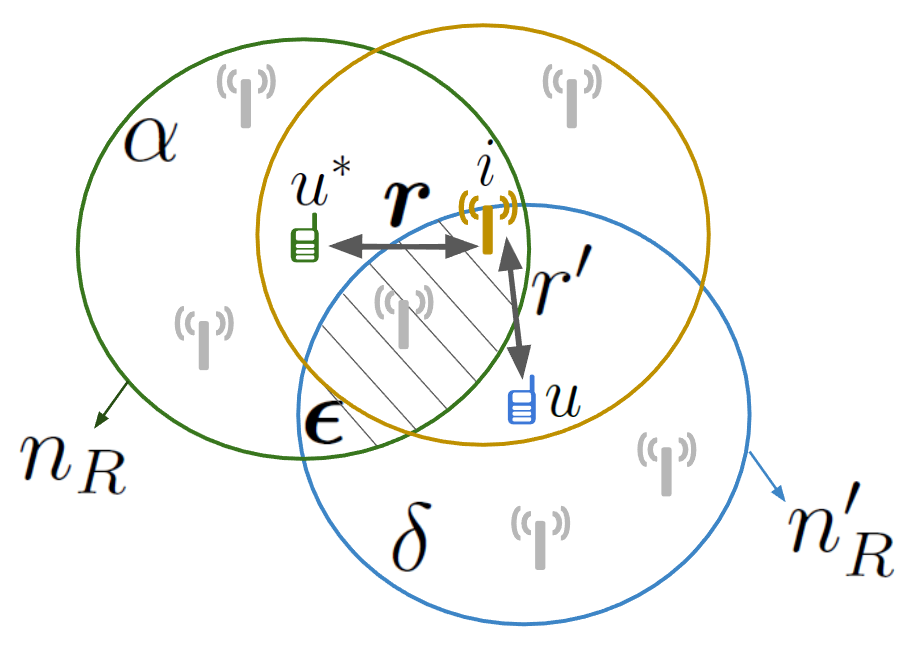}
         \caption{Intersection considered for $p(n_R'|n_R,r)$}
         \label{hyp_spatial_interf_2}
     \end{subfigure}
        \caption{Intersections considered to compute the conditional probabilities.}
        \label{hyp_spatial_interf}
\end{figure}

These two probabilities can be computed using a geometrical reasoning based on Figure \ref{hyp_spatial_interf}.
In Figure \ref{hyp_spatial_interf_1}, the intersection area $\beta$ of the two disks around $u^*$ and $i$ is given by \cite{circle_intersect}:
\begin{equation}
    \mathcal{A}(r) = 2r_1^{2} \text{acos}\Big(\dfrac{r}{2r_1} \Big) - \dfrac{r}{2}\sqrt{4r_1^2-r^2}.
\end{equation}

Nb: in order to obtain simpler forms, this expression can also be replaced by its linearization around zero, without loss of significant accuracy:

\begin{equation}
    \Tilde{\mathcal{A}}(r) = \pi r_1^2 \bigg[1 - \frac{2r}{\pi r_1} \bigg].
\end{equation}

Since the users are distributed using a PPP, the number of nodes in the zones $\alpha$, $\beta$ and $\gamma$ follow independent Poisson distributions. The conditional probability of $n_U'$ is hence given by
\begin{align}
    p(n_U'|n_U) &= \dfrac{p\bigg[\Big(n_U \; \text{in} \; (\alpha \cup \beta) \Big) \cap \Big(n_U' \; \text{in} \; (\beta \cup \gamma) \Big) \bigg]}{p\Big[n_U \; \text{in} \;  (\alpha \cup \beta) \Big]} \\
    &= \dfrac{\sum_{k=0}^{\min(n_U,n_U')} p\big[n_U - k \; \text{in} \;  \alpha \big]p\big[k \; \text{in} \;  \beta \big]p\big[n_U'-k \; \text{in} \;  \gamma \big]}{p\big[n_U \; \text{in} \;  (\alpha \cup \beta) \big]}
\end{align}
where 
\begin{align*}
    p\big[n_U - k \; \text{in} \;  \alpha \big] &= \Big[\lambda_U \big(\pi r_1^2 - \mathcal{A}(r)\big)\Big]^{n_U-k} \exp\Big[- \lambda_U \big(\pi r_1^2 - \mathcal{A}(r)\big)\Big] \Big[(n_U-k)!\Big]^{-1}\\
    p\big[k \; \text{in} \;  \beta \big] &= \big[\lambda_U \mathcal{A}(r)\big]^{k} \exp\big[- \lambda_U \mathcal{A}(r)\big] \big[k!\big]^{-1} \\
    p\big[n_U'-k \; \text{in} \;  \gamma \big] &= \Big[\lambda_U \big(\pi r_1^2 - \mathcal{A}(r)\big) \Big]^{n_U'-k} \exp\Big[- \lambda_U \big(\pi r_1^2 - \mathcal{A}(r)\big)\Big] \Big[(n_U'-k)!\Big]^{-1}\\
    p\big[n_U \; \text{in} \;  (\alpha \cup \beta) \big] &= \big(\lambda_U \pi r_1^2 \big)^{n_U} \exp\big(- \lambda_U \pi r_1^2\big) \big(n_U! \big)^{-1}.
\end{align*}

Using the same reasoning, the probability of $n_R'$ is hence given by
\begin{align}
    \label{pp}
    p(n_R'|n_R) &= \dfrac{p\bigg[\Big(n_R \; \text{in} \; (\alpha \cup \epsilon) \Big) \cap \Big(n_R' \; \text{in} \; (\epsilon \cup \delta) \Big) \bigg]}{p\Big[n_R \; \text{in} \;  (\alpha \cup \epsilon) \Big]} \\
    &= \dfrac{\sum_{k=0}^{\min(n_R,n_R')} p\big[n_R - k \; \text{in} \;  \alpha \big]p\big[k \; \text{in} \;  \epsilon \big]p\big[n_R'-k \; \text{in} \;  \delta \big]}{p\big[n_R \; \text{in} \;  (\alpha \cup \epsilon) \big]}
\end{align}

\noindent
The probabilities involved in this last expression require to compute the area of the intersection $\epsilon$ in Figure \ref{hyp_spatial_interf_2}. This area depends on the distance between the green and blue circle, given by $r'' = \big(r^2+r'^2-2rr'\cos(\angle r r')\big)^{1/2}$. The resulting intersection should be average over the distribution of $r'$, which does not easily lead to tractable expressions. In order to work with a simple expression, an analytical expression $\mathcal{S}(r)$ only function of variable $r$. In addition, we consider a linear expression for the relative area of intersection: $ \mathcal{S}(r)/(\pi r_1^2) = \chi + \zeta r$, whose coefficients are determined by considering the average case $r'=r_1/2$ and $\angle r r' = \pi /2$. The resulting values are given by 
\begin{align}
    \chi = \bigg(2\text{acos}\Big(\frac{\sqrt{5}}{4}\Big) - \frac{5}{4} \sqrt{4-\frac{1}{4}}\bigg) && \zeta = \frac{1}{r_1}\bigg(\chi - 2\text{acos}\Big(\frac{1}{4}\Big) + \frac{1}{4} \sqrt{4-\frac{1}{4}}\bigg).
\end{align}

The probabilities present in (\ref{pp}) can in that case be expressed as 
\begin{align*}
    p\big[n_R - k \; \text{in} \;  \alpha \big] &= \Big[\lambda_R \big(\pi r_1^2 - \mathcal{S}(r)\big) \Big]^{n_R-k} \exp\Big[- \lambda_R \big(\pi r_1^2 - \mathcal{S}(r)\big)\Big] \Big[(n_R-k)!\Big]^{-1}\\
    p\big[k \; \text{in} \;  \epsilon \big] &= \Big[\lambda_R\mathcal{S}(r)\Big]^{k} \exp\Big[- \lambda_R \mathcal{S}(r) \Big] \Big[(k)!\Big]^{-1}\\
    p\big[n_R'-k \; \text{in} \;  \delta \big] &= \Big[\lambda_R\big(\pi r_1^2 - \mathcal{S}(r)\big)\Big]^{n_R'-k} \exp\Big[- \lambda_R \big(\pi r_1^2 - \mathcal{S}(r)\big)\Big] \Big[(n_R'-k)!\Big]^{-1}\\
    p\big[n_R \; \text{in} \;  (\alpha \cup \epsilon) \big] &= \big(\lambda_U \pi r_1^2 \big)^{n_R} \exp\big(- \lambda_R \pi r_1^2\big) \big(n_R! \big)^{-1}.
\end{align*}

\section{Proof of Proposition \ref{corr2}}
\label{proof_corr2}

In this proof, equations \eqref{uncondi}, \eqref{phi_S}, \eqref{phi_T} and \eqref{SandT} are successively derived.

\textbf{Equation (\ref{uncondi})}: conditioned on $n_U$ and $n_R$, the characteristic function of $P_{S_{\eta}}$ is given by
\begin{align*}
    \phi_{P_{S_{\eta}}}(t|n_R,n_U) &= \mathbb{E}\Big[\exp(jtP_{S_{\eta}}) \Big]\\
    &\stackrel{(a)}{=}  \mathbb{E}\Bigg[\exp\bigg\{jt\bigg(\sum_{i=1}^{n_R} P_{S,i} + \eta \sum_{i=1}^{n_R} P_{I_1,i} \bigg) \bigg\} \Bigg] \\
    &= \mathbb{E}\Bigg[ \prod_{i=1}^{n_R} \exp\Big(jtP_{S,i}\Big) \exp\Big(jt\eta P_{I_1,i}\Big) \Bigg]\\
    &\stackrel{(b)}{=} \Bigg(\mathbb{E}\bigg[\exp\Big(jtP_{S,i}\Big) \exp\Big(jt\eta P_{I_1,i}\Big) \bigg] \Bigg)^{n_R} \\
    &\stackrel{(c)}{=} \Bigg(\mathbb{E}_r\bigg[\phi_{T}(t|n_R,n_U,r)\phi_{V}(\eta t|n_R,n_U,r)\bigg] \Bigg)^{n_R} \\
    &\stackrel{(d)}{=} \Bigg[\int_{r_0}^{r_1} \phi_{T}(t|n_R,n_U,r)\phi_{V}(\eta t|n_R,n_U,r) \frac{2r}{r_1^2-r_0^2} dr\Bigg]^{n_R}.
\end{align*}

In these developments, (a) is obtained by decomposing $P_{S_{\eta}}$ into the useful and interference powers provided by each RRH $i \in \mathcal{C}_{u^*}$. (b) comes from the assumed independence between the different RRHs under conditioning on $n_U$ and $n_R$. (c) is obtained by defining $\phi_T(\cdot)$ and $\phi_V(\cdot)$ as the characteristic functions of powers $P_{S,i}$ and $P_{I_1,i}$ coming from an arbitrary RRH $i$. (d) is derived by averaging over the pdf of the distance $r_{iu^*}$ between the RRH and the user.

\textbf{Equation (\ref{phi_S})}: by replacing $\mathbf{w}_{iu^*}$ by its definition in (\ref{eqn_power}), we obtain 
\begin{equation}
\label{simple_form_useful}
    P_{S,i} = P_t|\mathbf{h}_{iu^*}|^2 \kappa^{-1}r_{iu^*}^{-\alpha}.
\end{equation}

This right member of (\ref{simple_form_useful}) follows a Gamma distribution of shape $M$ and scale $P_t\kappa^{-1}r_{iu^*}^{-\alpha}$, which results in $\phi_T(\cdot)$ given by (\ref{phi_S}).

\textbf{Equation (\ref{phi_T})}: $\phi_V(\cdot)$ is the characteristic function of the interference power coming from one RRH $i \in \mathcal{C}_{u^*}$. This interference power is given by
\begin{equation*}
 P_{I_1,i} =  \sum\limits_{\substack{u \in \mathcal{B}_i \\ u \neq u^*}} \;  P_t Z^{(i,u)}_{|\mathcal{C}_{u}|} \kappa^{-1} r^{-\alpha}_{iu^*}
\end{equation*}
On the basis of this definition, one has
\begin{align}
    \phi_{V}(t|n_R,n_U,r) &= \mathbb{E}\Big[\exp(jt P_{I_1,i}) \Big]\\
    &=  \mathbb{E}\Bigg[\exp\Bigg\{jt \Bigg( \sum\limits_{\substack{u \in \mathcal{B}_i \\ u \neq u^*}} \;  P_t Z^{(i,u)}_{|\mathcal{C}_{u}|} \kappa^{-1} r^{-\alpha}_{iu^*} \Bigg)\Bigg\} \Bigg] \\
    &\stackrel{(a)}{=}  \sum_{n_U'=0}^{\infty} p(n_U'|n_U,r) \; \mathbb{E}\Bigg[\exp\Bigg\{jt \Bigg( \sum\limits_{u=1}^{n_U'} \;  P_t Z^{(i,u)}_{|\mathcal{C}_{u}|} \kappa^{-1} r^{-\alpha} \Bigg)\Bigg\} \Bigg] \\
    &\stackrel{(b)}{=} \sum_{n_U'=0}^{\infty} p(n_U'|n_U,r) \; \Bigg( \mathbb{E}\Bigg[\exp\Bigg\{jt P_t Z^{(i,u)}_{|\mathcal{C}_{u}|} \kappa^{-1} r^{-\alpha}\Bigg\} \Bigg] \Bigg)^{n_U'} \\
    &\stackrel{(c)}{=} \sum_{n_U'=0}^{\infty} p(n_U'|n_U,r) \; \Bigg(\sum_{n_R' = 1}^{\infty} p(n_R'|n_R,r) \; \mathbb{E}\Bigg[e^{jt P_t Z^{(i,u)}_{n_R'} \kappa^{-1} r^{-\alpha}} \Bigg] \Bigg)^{n_U'} \\
    &\stackrel{(d)}{=} \sum_{n_U'=0}^{\infty} p(n_U'|n_U,r) \Bigg[\sum_{n_R' = 1}^{\infty} p(n_R'|n_R,r) \; \big(1-jtP_t \kappa^{-1}r^{-\alpha}s_{n_R'}\big)^{-k_{n_R'}} \Bigg]^{n_U'}
\end{align}

In these developments, (a) comes from the averaging over $n_U' = |\mathcal{B}_i| -1$, the number of UEs different from $u^*$ served by RRH $i$. (b) comes from the assumed independence between the interference terms associated to the $n_U'$ users under conditioning on $n_U$ and $n_R$. (c) is obtained via averaging over the $n_R'$ RRHs serving an arbitrary $u$ among the $n_U'$ UEs. (d) is finally derived by definition of the characteristic function of the Gamma random variable $Z^{(i,u)}_{n_R'}$.

\textbf{Equation (\ref{SandT})}: this final characteristic function is finally obtained by averaging over the (Poisson) distributions of $n_R$ and $n_U$.

\section{Proof of Proposition \ref{mv_inside}}
\label{proof_mean1}
The proof consists in computing the expectation of $P_{S_{\eta}}$ and $P_{S_{\eta}}^2$ for $\eta=1$. The steps of the developments rely on the same hypotheses as the proof of Proposition \ref{corr2}. The mean can be expressed as follows: 
\begin{align}
    m_{S+P_{I_1}} &= \mathbb{E}\Bigg[ \sum_{i \in \mathcal{C}_{u^*}} P_t |\mathbf{h}_{iu^*}|^2 \kappa^{-1} r^{-\alpha}_{iu^*} + \sum_{i \in \mathcal{C}_{u^*}} \; \sum_{\substack{u \in \mathcal{B}_i \\ u \neq u^*}} \; P_t  Z^{(i,u)}_{|\mathcal{C}_{u}|} \kappa^{-1} r^{-\alpha}_{iu^*} \Bigg] \\
    &\stackrel{(a)}{=} P_t \kappa^{-1}  \mathbb{E}\Bigg[ \sum_{i \in \mathcal{C}_{u^*}} \Bigg(|\mathbf{h}_{iu^*}|^2 + \sum_{\substack{u \in \mathcal{B}_i \\ u \neq u^*}} \;  Z^{(i,u)}_{|\mathcal{C}_{u}|}\Bigg) r^{-\alpha}_{iu^*} \Bigg] \\
    &\stackrel{(b)}{=} P_t \kappa^{-1} \sum_{n_U = 0}^{\infty}\sum_{n_R = 0}^{\infty} p_R(n_R)p_U(n_U) \underbrace{\mathbb{E}\Bigg[ \sum_{i =1}^{n_R} \Bigg(|\mathbf{h}_{iu^*}|^2 + \sum_{\substack{u \in \mathcal{B}_i \\ u \neq u^*}} \;  Z^{(i,u)}_{|\mathcal{C}_{u}|}\Bigg) r^{-\alpha}_{iu^*} \Bigg| n_R,n_U\Bigg] }_{L(n_U,n_R)} 
\end{align}
where (a) is obtained by grouping terms and (b) by conditioning on variables $n_U$ and $n_R$, as performed in Appendix \ref{proof_corr2}. The term $L(n_U,n_R)$ can then be written as
\begin{align}
L(n_U,n_R) &\stackrel{(a)}{=} n_R \mathbb{E}\Bigg[ \Bigg(|\mathbf{h}_{iu^*}|^2 + \sum_{\substack{u \in \mathcal{B}_i \\ u \neq u^*}} \;  Z^{(i,u)}_{|\mathcal{C}_{u}|}\Bigg) r^{-\alpha}_{iu^*} \Bigg| n_R,n_U\Bigg] \\
&\stackrel{(b)}{=} n_R \int_{r_0}^{r_1} \mathbb{E}\Bigg[ \Bigg(|\mathbf{h}_{iu^*}|^2 + \sum_{\substack{u \in \mathcal{B}_i \\ u \neq u^*}} \;  Z^{(i,u)}_{|\mathcal{C}_{u}|}\Bigg)  \Bigg| r, n_R,n_U\Bigg] r^{-\alpha} f_R(r) dr \\
&\stackrel{(c)}{=} n_R \int_{r_0}^{r_1} \Bigg(M + \underbrace{\mathbb{E}\Bigg[\sum_{\substack{u \in \mathcal{B}_i \\ u \neq u^*}} \;  Z^{(i,u)}_{|\mathcal{C}_{u}|}  \Bigg| r, n_R,n_U\Bigg]}_{\xi} \Bigg) r^{-\alpha} f_R(r) dr
\end{align}
where (a) is obtained by linearity. The rest of the developments are hence performed for an arbitrary RRH $i$ among the $n_R$ RRHs serving $u^*$. (b) is obtained by developing the averaging with respect to the distance $r_{iu^*}$. We denote the probability density function of this distance by $f_R(r) = 2r/(r_1^2-r_0^2)$. The remaining term $\xi$ is then computed by expanding then averaging over variables $n_R'$ and $n_U'$ introduced in section III.B. To this purpose, the conditional probabilities computed in Assumption \ref{pre_corr2} are employed. One obtains:
\begin{align}
    \xi &\stackrel{(a)}{=} \sum_{n_U' = 0}^{\infty} p(n_U'|n_U,r) \mathbb{E}\Bigg[\sum_{u=1}^{n_U'} Z^{(i,u)}_{|\mathcal{C}_{u}|} \Bigg| r, n_R,n_U\Bigg] \\
    &\stackrel{(b)}{=} \sum_{n_U' = 0}^{\infty} p(n_U'|n_U,r) n_U' \mathbb{E}\Bigg[Z^{(i,u)}_{|\mathcal{C}_{u}|} \Bigg| r, n_R,n_U,n_U'\Bigg] \\
    &\stackrel{(c)}{=} \sum_{n_U' = 0}^{\infty} p(n_U'|n_U,r) n_U' \sum_{n_R' = 0}^{\infty} p(n_R'|n_R,r) \mathbb{E}\Bigg[Z^{(i,u)}_{n_R'} \Bigg| r, n_R,n_U,n_R'\Bigg] \\
    &\stackrel{(d)}{=} \sum_{n_U' = 0}^{\infty} p(n_U'|n_U,r) n_U' \sum_{n_R' = 0}^{\infty} p(n_R'|n_R,r) n_R'^{-1}
\end{align}
where (a) is obtained by expanding the averaging with respect to $n_U' \triangleq |\mathcal{B}_i|$. (b) is derived by linearity. The rest of the developments are hence performed for one unique term of the summation, associated to an arbitrary user $u$ among the $n_U'$ UEs served by $i$. (c) is derived by developing the averaging with respect to $n_R' \triangleq |\mathcal{C}_u|$. 
In order to compute the variance, the computation of $\mathbb{E}[P_{S_{\eta}}^2]$ follows the same reasoning and is here omitted due to space limitations.
\section{Proof of Proposition \ref{outside}}
\label{proof_outside}
On the basis of the proposed decomposition, the interference $P_{I_2}$ can be expressed as
\begin{align}
    P_{I_2} = \sum_{n=1}^{\infty} \; \underbrace{\sum_{i \in \Tilde{\Psi}_{R,n}} \; \sum_{u=1}^{n} \; P_t  Z^{(i,u)}_{|\mathcal{C}_{u}|} \kappa^{-1} r^{-\alpha}_{iu^*}}_{P_{I_2,n}}.
\end{align}
where $P_{I_2,n}$ is the interference coming from $\Tilde{\Psi}_{R,n}$.
Since the subsets of the decomposition are independent, the characteristic function of $P_{I_2}$ can is given by
\begin{align}
    \phi_{P_{I_2}}(t) = \prod_{n=1}^{\infty} \phi_{P_{I_2,n}}(t).
\end{align}
where $\phi_{P_{I_2,n}}(t)$ is the characteristic function of the interference coming from the set $\Tilde{\Psi}_{R,n}$. This function is by definition given by
\begin{align}
\phi_{P_{I_2,n}}(t) &= \mathbb{E} \Bigg\{ \prod_{i \in \Tilde{\Psi}_{R,n}} \exp\bigg[jt\sum_{u=1}^{n} \; P_t  Z^{(i,u)}_{|\mathcal{C}_{u}|} \kappa^{-1} r^{-\alpha}_{iu^*} \bigg] \Bigg\}
\end{align}
Introducing the slack variable $Y_{i,n} = \sum_{u=1}^{n} \;  Z^{(i,j)}_{|\mathcal{C}_{u}|}$, this expression can be rewritten as
\begin{align}
    \phi_{P_{I_2,n}}(t) &= \mathbb{E}_{\Tilde{\Psi}_{R,n}} \Bigg\{ \prod_{i \in \Tilde{\Psi}_{R,n}} \mathbb{E}_{Y_{i,n}}\Big[ \exp \big(jtP_t\kappa^{-1}r^{-\alpha}_{iu^*}Y_{i,n} \big) \Big] \Bigg\} \\  \
    &\stackrel{(a)}{=} \mathbb{E}_{\Tilde{\Psi}_{R,n}} \Bigg\{ \prod_{i \in \Tilde{\Psi}_{R,n}} \phi_{Y_{i,n}}(P_t \kappa^{-1} r^{-\alpha}_{iu^*} t) \Bigg\} \\ \
    &\stackrel{(b)}{=} \exp\Bigg\{-2\pi \lambda_{R,n}\int_{r_1}^{\infty} \Big[1 - \phi_{Y_{i,n}}(P_t \kappa^{-1} r^{-\alpha} t) \Big] rdr \Bigg\} 
\end{align}
where (a) comes from the definition of the characteristic function and (b) is obtained using the probability generating functional (PGFL) of a PPP \cite{Baccelli}. 

Finally, the characteristic function of $Y_{i,n}$ is given by 
\begin{align}
\phi_{Y_{i,n}}(t) &= \mathbb{E}\bigg[\exp\bigg(jt \sum_{u=1}^{n} \;  Z^{(i,u)}_{|\mathcal{C}_{u}|} \bigg)\bigg] \\
 &\stackrel{(a)}{=} \Bigg(\mathbb{E}\bigg[\exp\Big(jt Z^{(i,u)}_{|\mathcal{C}_{u}|} \Big)\bigg]\Bigg)^n \\
 &\stackrel{(b)}{=} \Bigg(\mathbb{E}_{|\mathcal{C}_{u}|}\bigg[\mathbb{E}_{Z^{(i,u)}_{|\mathcal{C}_{u}|}} \Big[ \exp\big(jt Z^{(i,u)}_{|\mathcal{C}_{u}|} \big) \big| \; |\mathcal{C}_{u}| \; \Big]  \bigg]\Bigg)^n \\
 &\stackrel{(c)}{=} \Bigg(\mathbb{E}_{|\mathcal{C}_{u}|} \bigg[ \big(1 - jts_{|\mathcal{C}_{u}|} \big)^{k_{|\mathcal{C}_{u}|}} \bigg]\Bigg)^n \\
 &\stackrel{(d)}{=} \Bigg(\sum_{m=1}^{\infty} \Tilde{p}_R(m) (1-jts_m)^{k_m} \Bigg)^n
\end{align}

where (a) comes from the fact that the variables $Z^{(i,j)}_{|\mathcal{C}_{j}|}$ are i.i.d. (b) is obtained by decomposing the expectation and (c) is derived using the characteristic function of the gamma random variable $Z^{(i,u)}_{|\mathcal{C}_{u}|}$ (cfr. Appendix \ref{proof_CT+moment_matching}). (d) is obtained by averaging over the possible values of $|\mathcal{C}_{u}|$, whose pmf is given by $\Tilde{p}(m)$. Note that the truncated Poisson distribution of \eqref{truncated} is here used since the each considered user is served at least by one RRH for $n>1$.

\section{Proof of Proposition \ref{GP_corr}}
\label{proof_GP_corr}

The data rate distribution can be expressed as
\begin{align}
    \mathcal{P}_r(\theta|n_U,n_R) &\triangleq \mathbb{P}\Big[\log_2 \big(1+ P_S/(P_I+N_0)\big) >\theta \Big] \\
    &=\mathbb{P}\Big[P_S - \underbrace{(2^{\theta}-1)}_{\Tilde{\theta}} P_I - \underbrace{(2^{\theta}-1)}_{\Tilde{\theta}} N_0 > 0 \Big]\\
    &= \mathbb{P}\Big[\underbrace{P_S - \Tilde{\theta} P_{I_1}}_{P_{S_{-\Tilde{\theta}}}} - \Tilde{\theta} P_{I_2} \Tilde{\theta} N_0 > 0 \Big] 
\end{align}
where the last equality comes from the decomposition of the interference into $P_{I_1}$ and $P_{I_2}$. The final result \eqref{coverage_eq} is finally obtained by defining the auxiliary variable $P_{S_{-\Tilde{\theta}}}$ (equal to $P_{S_{\eta}}$ defined in section III.B, for the particular case of $\eta = -\Tilde{\theta}$) and by applying Gil-Pelaez theorem \cite{GP}. The reasoning employed to derive the IPD distribution is identical. 

\section*{Acknowledgment}

This work was supported by F.R.S.-FNRS under the EOS program (EOS project 30452698).

\ifCLASSOPTIONcaptionsoff
  \newpage
\fi

\small

\bibliographystyle{IEEEtran}
\bibliography{References}

\end{document}